\documentclass{ws-ijmpb}

\usepackage[LGR,T1]{fontenc}
\usepackage{amsmath}
\usepackage{amssymb}
\usepackage{graphicx}

\makeatletter

\DeclareRobustCommand{\greektext}{%
  \fontencoding{LGR}\selectfont\def\encodingdefault{LGR}}
\DeclareRobustCommand{\textgreek}[1]{\leavevmode{\greektext #1}}
\DeclareFontEncoding{LGR}{}{}
\DeclareTextSymbol{\~}{LGR}{126}
\providecommand{\tabularnewline}{\\}
\newcommand{\lyxdot}{.}

\@ifundefined{textcolor}{}
{%
 \definecolor{BLACK}{gray}{0}
 \definecolor{WHITE}{gray}{1}
 \definecolor{RED}{rgb}{1,0,0}
 \definecolor{GREEN}{rgb}{0,1,0}
 \definecolor{BLUE}{rgb}{0,0,1}
 \definecolor{CYAN}{cmyk}{1,0,0,0}
 \definecolor{MAGENTA}{cmyk}{0,1,0,0}
 \definecolor{YELLOW}{cmyk}{0,0,1,0}
}

\newcommand{\tr}{\mathrm{tr}}
\newcommand{\hc}{\mathrm{h.c.}}
\newcommand{\1}{\leavevmode{\rm 1\ifmmode\mkern  -4.8mu\else\kern -.3em\fi I}}

\makeatother

\begin{document}

\title{Theory of temporal fluctuations in isolated quantum systems}

\author{Lorenzo Campos Venuti and Paolo Zanardi}

\address{Department of Physics and Astronomy \& Center for Quantum Information
Science \& Technology,\\
 University of Southern California, Los Angeles, CA 90089-0484, USA}
 
 \maketitle
 
\begin{abstract}
When an isolated quantum system is driven out of equilibrium, expectation
values of general observables start oscillating in time. This article
reviews the general theory of such \emph{temporal fluctuations}. We
first survey some results on the strength of such temporal fluctuations.
For example temporal fluctuations are exponentially small in the system's
volume for generic systems whereas they fall-off algebraically in
integrable systems. We then concentrate on the the so-called quench
scenario where the system is driven out-of-equilibrium under the application
of a sudden perturbation. For sufficiently small perturbations, temporal
fluctuations of physical observables can be characterized in full
generality and can be used as an effective tool to probe quantum criticality
of the underlying model. In the off-critical region the distribution
becomes Gaussian. Close to criticality the distribution becomes a
universal function uniquely characterized by a single critical exponent,
that we compute explicitly. This contrasts standard equilibrium quantum
fluctuations for which the critical distribution depends on a numerable
set of critical coefficients and is known only for limited examples.
The possibility of using temporal fluctuations to determine pseudo-critical
boundaries in optical lattice experiments is further reviewed. 
\end{abstract}


\maketitle

\section{Introduction}

In the last few years we have witnessed a strong revival of interest
in foundational issues of quantum statistical mechanics \cite{lloyd__1988,goldstein_canonical_2006,popescu_entanglement_2006,rigol_relaxation_2007,reimann_foundation_2008,rigol_thermalization_2008,goldstein_normal_2010,polkovnikov_colloquium:_2011,yurovsky_dynamics_2011}.
The central questions of this field go back to the time of Boltzmann:
how do closed quantum systems out-of-equilibrium eventually equilibrate?
Under which circumstances can we justify the amazing effectiveness
of statistical ensembles in predicting the equilibrium properties
of macroscopic observables in physical systems? Whereas understanding
of thermal equilibrium is possible in the framework of quantum statistical
mechanics we still do not know how thermal equilibrium is reached
following the microscopic dynamical laws. What are the time-scales
associated to thermalization, and what are the conditions leading
to it? To date, even such simple questions lack a precise answer.
Thanks to current advances in experimental techniques, isolated quantum
systems are now routinely observed and the emergence of equilibrium
can be experimentally put to test (see e.g.~\cite{greiner_collapse_2002,greiner_quantum_2002,kinoshita_quantum_2006}).
Motivated by such experiments there has been a tremendous effort in
understanding how thermal equilibrium is reached. However, despite
the sheer amount of results which have accumulated, the precise conditions
which lead to thermalization remain to a large extent, unknown.

Understanding how thermalization is achieved amounts to understand
two conceptually separate physical processes. On the one hand an ``equilibrium''
state emerges out of the dynamical evolution. This equilibration process
is potentially much more general than thermalization itself. On the
other hand one asks how and why this equilibrium state has the familiar
thermal form expected from statistical mechanics, i.e.~is a microcanonical,
Gibbs, or grandcanonical thermal state depending on the appropriate
ensemble. Indeed a great effort has been put in understanding the
latter question. One of the central mechanisms that have been proposed
is the so called Eigenstate Thermalization Hypothesis \cite{deutsch_quantum_1991,srednicki_chaos_1994,srednicki_approach_1999,rigol_thermalization_2008}
(see also \cite{jensen_statistical_1985} and the recent \cite{gogolin_equilibration_2015}
for a comprehensive review). In this article however, we will concentrate
on the equilibration process alone and will not be concerned if equilibration
is of thermal type. 

The first question we address is then, in which sense an equilibrium
state can emerge out of a unitary dynamics? As it will be discussed
in Sec.~\ref{sec:Equilibration}, equilibration in finite systems
(or more generally in systems with discrete spectrum) must be formulated
in a probabilistic fashion. In an out-of-equilibrium system, the expectation
value of a quantum observable $A$, becomes a time-dependent quantity
$\mathsf{A}(t):=\langle A(t)\rangle$ oscillating around an average
value. In the spirit of the ergodic theory, one must introduce a large
\emph{observation} time window $[0,T]$. The \emph{time-fluctuations}
of $\langle A(t)\rangle$ are conveniently characterized by a temporal
probability distribution function $P_{A}(a)$ where the full time
statistics of $\langle A(t)\rangle$ is encoded. $P_{A}(a)da$ is
the probability that $\langle A(t)\rangle\in[a,a+da]$ for $t$ in
the observation window. Denoting with $\overline{f}=T^{-1}\int_{0}^{T}f(t)dt$
the time average operation, the equality $\overline{\mathsf{A}}:=\overline{\langle A(t)\rangle}=\tr(A\overline{\rho})$
shows that $\overline{\rho}$ plays the role of equilibrium state.
Observables expectation values, $\mathsf{A}(t)$, oscillate around
their averages $\overline{\mathsf{A}}$ with certain fluctuations
encoded in the distributions $P_{A}(a)$. We can say that equilibration
is reached if such fluctuations are small in some sense. In essence,
equilibration in closed, finite, quantum systems, corresponds to concentration
results (i.e~the \textquotedbl{}peakedness\textquotedbl{}) of these
probability distributions $P_{A}(a)$. The purpose of these article
is to illustrate that a wealth of physical information can be revealed
from the study of the full time statistics. In well defined scenarios,
the analysis of the full time statistics allows one to spot the precise
location of the underlying quantum critical points \cite{campos_venuti_universality_2010,campos_venuti_universal_2014},
or the integrable-non-integrable transition\cite{campos_venuti_gaussian_2013}.
In general, the full temporal statistics of a given observable is
an experimentally accessible quantity that encodes the physical data
in a convenient way. In the sequel we will use natural units throughout
in which $\hbar=k_{B}=1$.

\section{Equilibration in closed quantum systems\label{sec:Equilibration}}

Let us now lay down the setup of the problem in a general and formal
way. The closed-system dynamics is described by the time-evolution
operator $U(t)=e^{-itH}$. Let also $H=\sum_{n}E_{n}\Pi_{n}$ be the
spectral resolution of the system Hamiltonian ($\Pi_{n}$'s spectral
projections). Closed quantum systems evolve unitarily and therefore
cannot converge in a strong sense to an equilibrium state starting
out from a generic pure state. Indeed, call $\overline{\rho}$ the
equilibrium state. $\overline{\rho}$ must be a fixed point of the
dynamics, i.e.~$U(t)\overline{\rho}U(t)^{\dagger}=\overline{\rho}$.
But then, calling $\rho(t)=U(t)\rho_{0}U(t)^{\dagger}$, one has $\left\Vert \rho(t)-\overline{\rho}\right\Vert =\left\Vert \rho(t)-U(t)\overline{\rho}U(t)^{\dagger}\right\Vert =\left\Vert \rho_{0}-\overline{\rho}\right\Vert =\mathrm{const}.$
In words, the distance between the state at at time $t$ and the equilibrium
state is constant. Despite of this fact, for sufficiently large system
sizes one may observe \emph{temporal typicality}. Namely, for the
overwhelming majority of the time instants, the statistics of observables
are practically indistinguishable from an effective equilibrium state.
In this sense, equilibration in isolated quantum systems emerges in
a probabilistic fashion. 

One may then wonder whether a weaker form of convergence can be achieved
for $t\to\infty$. Let us therefore consider the expectation value
of an observable $\mathcal{\mathsf{A}}(t):={\rm tr[\rho(t)A]}$, using
the the spectral resolution of the Hamiltonian one finds a time-independent
contribution to $\mathcal{\mathsf{A}}(t)$, i.e.~$\mathsf{A}_{\infty}:=\sum_{n}{\rm {tr}}(\Pi_{n}\rho_{0}\Pi_{n}A)$
plus a time-dependent one $\tilde{\mathcal{\mathsf{A}}}(t)$. The
point is to understand whether this latter term admits a limit for
$t\rightarrow\infty$ (see also Ref.~\refcite{ziraldo_relaxation_2012}).
In finite dimensions, it is easy to see from the discrete nature of
the spectrum of $H$ that $\mathsf{\tilde{A}}(t)$ is a quasi-periodic
function:%
\footnote{Mathematicians call such functions \emph{almost-periodic}, and their
properties have been studied extensively %
} \emph{the infinite time limit of $\mathsf{A}(t)$ does not exist}.
On the other hand in the infinite-dimensional case the spectrum of
$H$ can be continuous and in this case the infinite time limit can
exist (essentially thanks to the Riemann-Lebesgue lemma) in which
case one has $\lim_{t\to\infty}\mathsf{A}(t)=\overline{\mathsf{A}(t)}=\mathsf{A}_{\infty}$,
where $\overline{\mathsf{A}(t)}:=\lim_{T\to\infty}T^{-1}\int_{0}^{T}\mathsf{A}(t)dt$
denotes the time-average over an infinite time interval.

There is a third form of convergence that one can consider here: \emph{the
convergence in probability}. In the following, we consider the above
defined $\mathsf{A}(t)$ as a random variable over the the interval
$[0,T]$ endowed with the uniform measure $dt/T$ with $T\rightarrow\infty$%
\footnote{The observation time $T$ is usually much larger than the typical
timescales of the system dynamics. Accordingly, here and in the following,
time averages are computed in the $T\rightarrow\infty$ limit if not
explicitly stated otherwise. %
}. Note that $\mathsf{A}(t)$ depends on the system size $L$. A compact
expression for the probability density of $\mathsf{A}$ is given by
$P_{A}(a):=\overline{\delta\left(a-\langle A\left(t\right)\rangle\right)}$
and encodes the full time statistics of $\mathsf{A}$. Thus $\int_{\Omega}P_{A}(a)d\alpha$
gives the probability that $\langle A\left(t\right)\rangle$ is in
$\Omega$ during the observation time $T.$ The equality $\mathsf{\overline{A}}=\overline{\tr\left[A\rho\left(t\right)\right]}=\tr\left[A\overline{\rho}\right]$
shows that the time-averaged state $\overline{\rho}$ plays the role
of the equilibrium density matrix. We say that $\mathsf{A}(t)$ converges
in probability to $\mathsf{A}_{\infty}$ if $\lim_{L\to\infty}P_{A}(a)=\delta(a-\mathsf{A}_{\infty})$,
in which case one must have $\mathsf{A}_{\infty}=\lim_{L\to\infty}\overline{\mathsf{A}}$.
At finite, fixed size, we say that an observable $A$ equilibrates
towards the mean $\overline{\mathsf{A}}$ if $\mathsf{A}(t)$ stays
close to $\overline{\mathsf{A}}$ for most of the times $t$ during
the observation interval $\left[0,T\right]$. Hence equilibration
of the observable $A$ is a concentration result of the distribution
of $\langle A\left(t\right)\rangle$.

\section{Temporal fluctuations}

In general the time signal $\mathsf{A}(t):=\langle A\left(t\right)\rangle=\tr(Ae^{-itH}\rho_{0}e^{itH})$
is a complicated function containing an overabundant amount of information.
For simplicity we stick here to the case where $\rho_{0}$ is a generic
\emph{pure }initial state, i.e.~$\rho_{0}=|\psi_{0}\rangle\langle\psi_{0}|$.
The distribution $P_{A}(a)$, instead, can be characterized by a much
smaller number of parameters (e.g.~mean, variance, some higher cumulants)
as a result of the high dimensionality of the system (the measure
concentration phenomenon). This allows a drastic simplification whereby physical properties are encoded in few parameters as opposed
to the $O\left(d^{2}\right)$ ($d$ Hilbert's space dimension) contained
in $\mathsf{A}(t)$.

Clearly the first possibility that comes to mind is that $P_{A}\left(a\right)$
be Gaussian for sufficiently large sizes, a situation that we refer
to as \emph{Gaussian equilibration}. Let $\Delta\mathsf{A}^{2}$ indicate
the temporal variance, i.e.~$\Delta\mathsf{A}^{2}=\overline{\mathsf{A}^{2}(t)}-\overline{\mathsf{A}(t)}^{2}$.
In the Gaussian equilibration scenario the relative fluctuations decay
as $\Delta\mathsf{A}/\overline{\mathsf{A}}\sim1/\sqrt{V}$ for increasing
system volume $V$%
\footnote{Here and throughout the paper, we indicate with ``volume'' the number
of elementary cells of the system, or the total volume normalized
to the single cell. As such it is a dimensionless number. %
} implying relatively large fluctuations for small sizes. Concerning
the variance $\Delta\mathsf{A}^{2}$, one can prove under some limiting
assumption (the so called non-resonant condition), the following simple
yet important result \cite{reimann_foundation_2008} 
\begin{equation}
\Delta\mathsf{A}^{2}\le\|A\|^{2}{\rm {tr}}\overline{\rho}^{2},\label{bound-varia}
\end{equation}
where $\|A\|=\sup_{\|\psi\|=1}\|A\psi\|$ is the norm of $A$%
\footnote{The bound (\ref{bound-varia}) can be slightly strengthened to $\Delta\mathsf{A}^{2}\le D(A)^{2}{\rm tr}\overline{\rho}^{2}$
where $D(A)=[\sup\sigma(A)-\inf\sigma(A)]/2$, and $\sigma(A)$ being
the spectrum of $A$. We won't be needing this slight generalization. %
}. The non-resonant condition is a condition on the degree of independence
of the energy levels. The precise statement is that from $E_{i}-E_{j}=E_{n}-E_{m}$
it follows either $i=j$ and $n=m$ or $i=n$ and $j=m$. This conditions
is notably violated for quasi-free systems and we will discuss its
consequences at length in the sequel. However it is believed to be
satisfied for generic, realistic models. Now it is possible to show
that, for \emph{generic} initial states and local Hamiltonian $H$,
the purity $\mathrm{tr}\left(\overline{\rho}^{2}\right)$ is exponentially
small in the system size \cite{campos_venuti_unitary_2010,campos_venuti_gaussian_2013}.
The precise condition is that $|\psi_{0}\rangle$ be sufficiently
clustering, meaning that, connected correlations of local observables
falls off e.g.~exponentially when two points are taken far apart.
This is known to be the case if the model is gapped \cite{nachtergaele_lieb-robinson_2006}.
The argument goes as follows. The first point is to notice that 
\[
\mathrm{tr}\left(\overline{\rho}^{2}\right)=\overline{\mathcal{L}(t)}
\]
where $\mathcal{L}(t)$ is the so called Loschmidt echo (LE) or survival
probability, 
\begin{equation}
\mathcal{L}(t):=\tr[\rho(t)\rho_{0}]=\tr[e^{-itH}\rho_{0}e^{itH}\rho_{0}].
\end{equation}
The Loschmidt echo arises in quite a few contexts in physics such
as quantum chaos \cite{jalabert_environment-independent_2001,karkuszewski_quantum_2002,casati_quantum_2006},
the theory of Fermi-edge singularity in the X-ray spectra of metals
\cite{schotte_tomonagas_1969,nozieres_singularities_1969} or the
physics of dephasing \cite{quan_decay_2006,rossini_decoherence_2007}.
Using a cumulant expansion the LE can be cast in the following way
\begin{eqnarray}
\mathcal{L}\left(t\right) & = & \exp2\sum_{n=1}^{\infty}\frac{\left(-t^{2}\right)^{n}}{\left(2n\right)!}\left\langle H^{2n}\right\rangle _{c},\label{eq:losch_cum}
\end{eqnarray}
where $\left\langle \cdot\right\rangle _{c}$ stands for the connected
average with respect to $\rho_{0}$. The above sum starts from $n=1$
because the zero order cumulant is zero: $\left\langle H^{0}\right\rangle _{c}=0$.
Assuming that initial state is clustering (e.g.~exponentially), and
given the fact that the Hamiltonian is a local, extensive, operator
(i.e.~$H=\sum_{x}h(x)$), all the cumulants are extensive: $\langle H^{n}\rangle_{c}\propto V$,
meaning that, for sufficiently large sizes $\mathcal{L}(t)\simeq\exp[g(t)V]$
(the function $g(t)$ must exist since $\mathcal{L}(t)$ is a positive
almost-periodic function). Taking the infinite time average one obtains
$\mathrm{tr}\left(\overline{\rho}^{2}\right)\le e^{-\eta V}$, with
$\eta$ positive constant. This in turn implies that $\Delta\mathsf{A}/\overline{\mathsf{A}}\le O\left(e^{-\mathrm{const.}\times V}\right)$
so that Gaussian equilibration cannot be the general scenario but
rather a stronger form of concentration must take place in generic
situations.

\begin{figure}
\begin{centering}
\includegraphics[width=6cm,height=6cm]{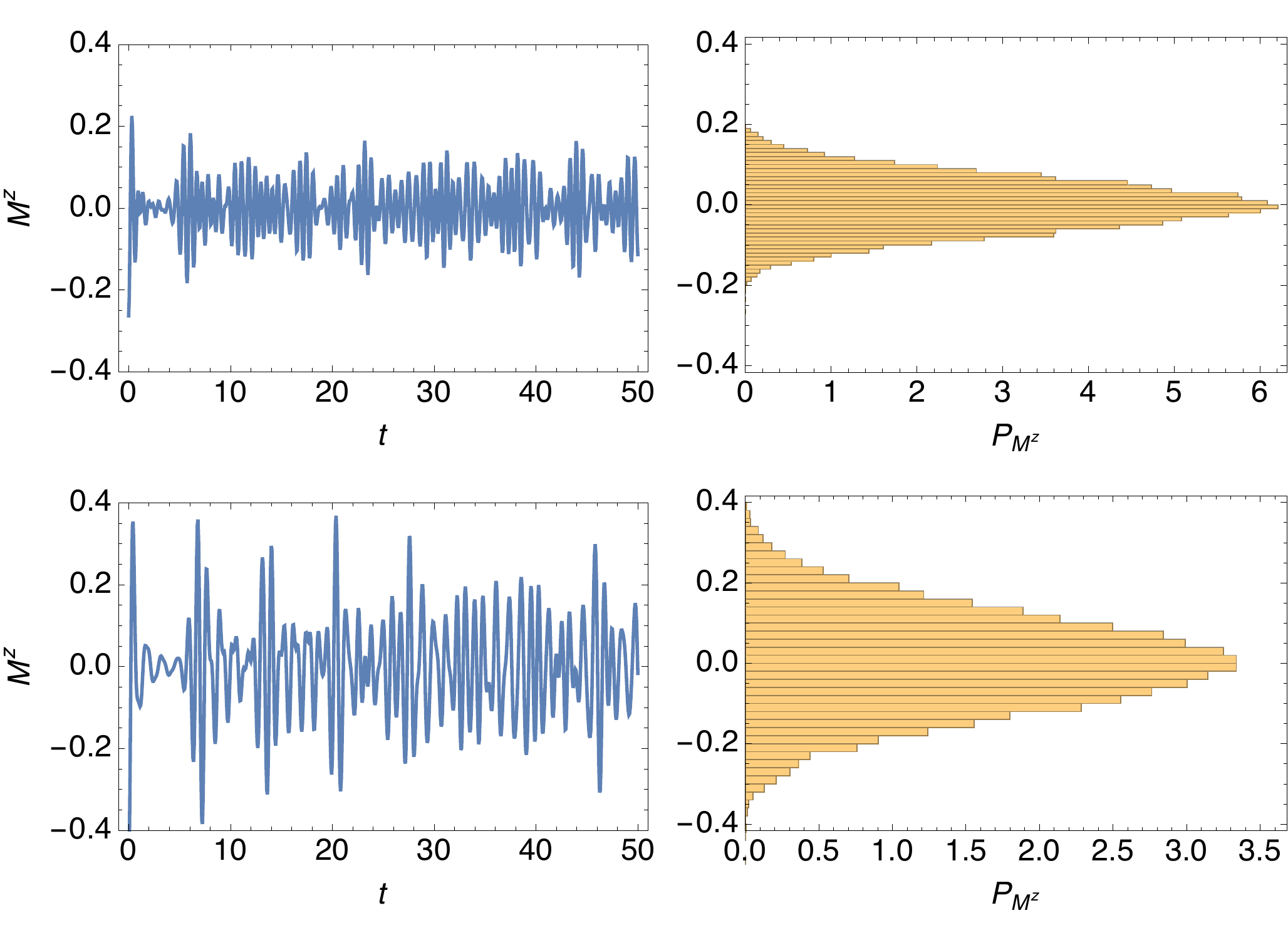}\hspace{4mm}\includegraphics[width=6cm,height=6cm]{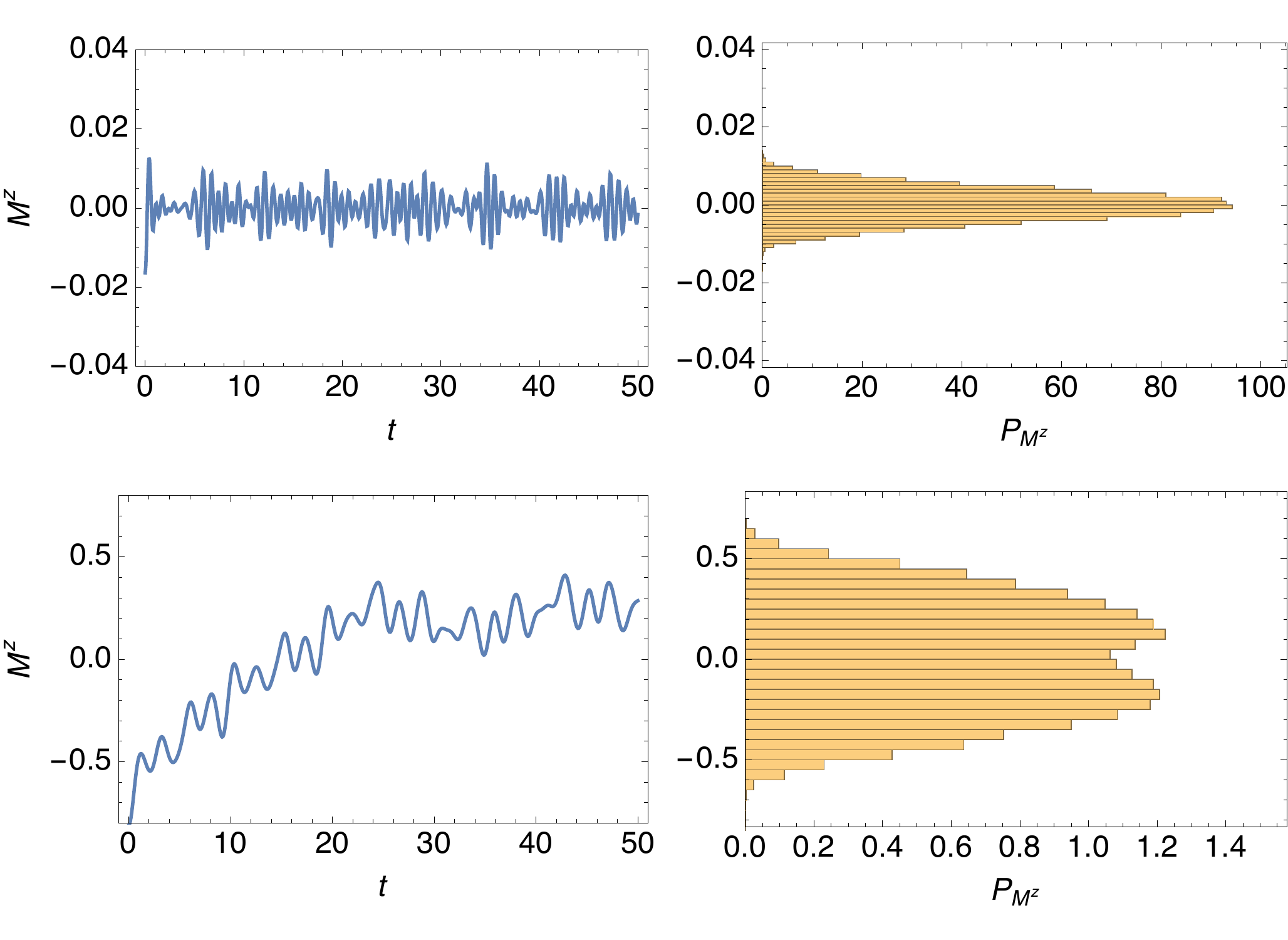}
\par\end{centering}
\protect\caption{Different behavior of temporal fluctuations illustrated by means of
quantum quenches on the Ising model with next nearest neighbor interaction
{[}Eq.~(\ref{eq:TAM}){]}. The observable is the total magnetization
$\mathsf{M}^{z}(t)=\sum_{j}\langle\sigma_{j}^{z}(t)\rangle$. Left
panels: ``large quenches''. Top left: generic quench in a non-integrable
system. Small fluctuations bounded by $\Delta\mathsf{M}^{z}=O(Le^{-\eta L})$,
bottom left: generic quench in an integrable system (and quadratic
observable). Gaussian distribution and fluctuations scaling as $\Delta\mathsf{M}^{z}\sim\sqrt{L}$.
The system is a chain with $L=12$ sites with open boundary conditions.
Parameters are $\kappa_{1}=\kappa_{2}=0.4$, $h_{1}=1.5$, $h_{2}=h_{1}+\delta h$,
$\delta h=0.5$ {[}top left{]}, $\kappa_{1}=\kappa_{2}=0.0$, $h_{1}=1.5$,
$h_{2}=h_{1}+\delta h$, $\delta h=0.5$ {[}bottom left{]}. Right
panels: ``small quenches''. Top right: small quench outside criticality,
Gaussian distribution, fluctuations scaling as $\Delta\mathsf{M}^{z}\sim\delta h\sqrt{L}$.
Bottom right: small quench close to criticality, bimodal distribution,
large fluctuations scaling as $\Delta\mathsf{M}^{z}\sim\delta\lambda L^{2\alpha}$,
with critical exponent $\alpha$ defined in section \ref{sub:Temporal-fluctuations-after}.
Parameters are, $L=12$, $\kappa_{1}=\kappa_{2}=0.4$, $h_{1}=1.5$,
$h_{2}=h_{1}+\delta h$, $\delta h=0.02$ {[}top right{]} $\kappa_{1}=\kappa_{2}=0.4$,
$h_{1}=0.218$, $h_{2}=h_{1}+\delta h$, $\delta h=0.02$ {[}bottom
right{]}. $h_{1},\kappa_{1}$ is a quantum critical point (see \protect\cite{beccaria_density-matrix_2006}).
\label{fig:summa_fig}}
\end{figure}

Nonetheless we have shown that indeed Gaussian equilibration is expected
in two important cases: i) quasi-free Fermi systems where both the
Hamiltonian and the observables are quadratic in Fermi operators,
for any, non-critical, generic initial state \cite{campos_venuti_gaussian_2013}
and, ii) small quench%
\footnote{The precise requirement is that of a weak perturbation meaning in
general $\delta\lambda/J\ll L^{-D}$, with $\delta\lambda$ quench
amplitude, $J$ energy scale of the unperturbed system, and $D$ spatial
dimension.%
} away from criticality\cite{campos_venuti_universality_2010,campos_venuti_exact_2011}.
Since Gaussian equilibration is a rather weak form of equilibration,
the result i) expresses in a precise and quantitative way the common
folklore that integrability leads to a poorer (or no) equilibration.
Likewise ii) can be explained noting that a small perturbation excites
relatively few quasi-particles and therefore results in poor equilibration.
In passing we note that Gaussian equilibration is not in contrast
with the bound (\ref{bound-varia}) in that quasi-free systems violate
the non-resonant condition which leads to Eq.~(\ref{bound-varia})
while for a small quench, the initial state is not generic and $\mathrm{tr}\overline{\rho}^{2}\approx1$
(in practice the constant $\alpha\to0$ in the small quench limit).

Another class of probability distributions $P_{A}(a)$ arises for
small quenches \emph{close to a quantum critical point}, a scenario
that will be described in more detail in Sec.~\ref{sub:Temporal-fluctuations-after}.
In this setting an even weaker form of equilibration takes place and
the full time statistics $P_{A}(a)$ for a generic observable $A$,
is predicted to have a universal bimodal shape.

These different forms of equilibration are illustrated by means of
numerical simulations on the transverse field Ising model with next
nearest neighbor interaction {[}see Sec.~\ref{sec:Temporal-fluctuations-for}
and Eq.~(\ref{eq:TAM}){]} in Figure \ref{fig:summa_fig}. In particular
we performed the so called quantum quench numerical experiment: the
system is initialized in the ground state of the Hamiltonian with
parameters $h_{1},\,\kappa_{1}$. One then suddenly changes the parameters
to $h_{2},\,\kappa_{2}$ and evolves the system with the resulting
Hamiltonian. The observable considered is the total transverse magnetization
$\mathsf{M}^{z}(t)=\sum_{j}\langle\sigma_{j}^{z}(t)\rangle$. In Fig.~\ref{fig:summa_fig}
we plot both the time series $\mathsf{M}^{z}(t)$ and its corresponding
probability distribution $P_{\mathsf{M}^{z}}$. A large quench is
a mean to initialize the system in a state which has little relation
with the evolution Hamiltonian. 

A large quench in a non-integrable model results in small fluctuations
scaling as $\Delta\mathsf{M}^{z}\sim Le^{-\eta L}$ {[}from Eq.~(\ref{bound-varia}){]}.
However, if the system is quasi-free (and the observable is quadratic)
the scaling becomes $\Delta\mathsf{M}^{z}\sim\sqrt{L}$, the Gaussian
equilibration scenario discussed in Sec.~\ref{sec:Temporal-fluctuations-for}.
One has a Gaussian distribution also for a small quench performed
in a gapped (non-critical) region of the phase diagram, with fluctuations
scaling as $\Delta\mathsf{M}^{z}\sim\delta h\sqrt{L}$ ($\delta h$
quench amplitude). Finally, if the small quench is performed close
to a quantum critical point, one obtains a bimodal distribution with
large fluctuations scaling as $\Delta\mathsf{M}^{z}\sim\delta hL^{2\alpha}$
{[}$\alpha$ critical exponent, see Section \ref{sub:Temporal-fluctuations-after}{]}.
For clarity these results are also summarized in Table \ref{tab:summa_table}. 

\begin{table}
\tbl{Scaling of the (temporal) standard deviation $\Delta\mathsf{A}$ of
an extensive observable for different scenarios. The asterisk refers
to a quadratic observable. $V$ is the system's volume, $L$ its linear
size, $\delta\lambda$ the quench amplitude and the critical exponent
$\alpha$ is defined in section \ref{sub:Temporal-fluctuations-after}.
The corresponding theories will be developed in sections \ref{sec:Temporal-fluctuations-for}
and \ref{sub:Temporal-fluctuations-after}. }
{\begin{tabular}{|c|c|c|c|} 
\hline 
\multicolumn{2}{|c|}{Large quenches} & \multicolumn{2}{c|}{Small quenches}\tabularnewline
\hline 
\hline 
Non-integrable & \hspace{5mm} $\le Ve^{-\eta V}$ \hspace{5mm} & Non-critical & \hspace{5mm} $\delta\lambda\sqrt{V}$ \hspace{5mm}\tabularnewline
\hline 
Quasi-free$^{\ast}$ & $\sqrt{V}$ & Quasi-critical & $\delta\lambda L^{\alpha}$\tabularnewline
\hline 
\end{tabular}}
\label{tab:summa_table}
%
%

\end{table}

\section{Temporal fluctuations for quasi-free Fermi systems\label{sec:Temporal-fluctuations-for}}

We now consider a situation where both the Hamiltonian and the observable
are quadratic in Fermi creation and annihilation operator%
\footnote{Throughout the paper we use interchangeably the terms quasi-free or
quadratic, for observables quadratic in Fermi operators. %
}. Note however that the initial state can be generic (i.e.~not necessarily
a Gaussian state). The Hamiltonian is $H=\sum_{x,y}c_{x}^{\dagger}M_{x,y}c_{y}=c^{\dagger}Mc$
(notation $c^{\dagger}=\left(c_{1}^{\dagger},\ldots,c_{V}^{\dagger}\right)$,
$V$ number of sites%
\footnote{For a $D$-dimensional lattices, a possibly better notation would
be through a $D$-dimensional spatial label $\boldsymbol{x}$. We
consider regular lattices for which the total number of points is
$V=O(L^{D})$ with $L$ some linear size. With this caveat, a multidimensional
generalization is straightforward. %
}), and the observable has the form $A=\sum_{x,y}c_{x}^{\dagger}a_{x,y}c_{y}=c^{\dagger}ac$.
We will assume that $\left\Vert a\right\Vert =O\left(1\right)$ %
\footnote{To achieve this, for example, in the translation invariant case, it
suffices to have $a_{x,y}=a(x-y)$ sufficiently fast decaying. %
} as this guarantees that the expectation values of $A$ scales at
most extensively with the system volume %
\footnote{\protect By diagonalizing $a$ and exploiting unitary invariance
of the operator norm, one finds $\langle A\rangle\le\|A\|=\|\sum_{\mu}\alpha_{\mu}c_{\mu}^{\dagger}c_{\mu}\|\le\sum_{\mu}|\alpha_{\mu}|\|c_{\mu}^{\dagger}c_{\mu}\|=\sum_{\mu}|\alpha_{\mu}|\le\|a\|L$.
Here the $\alpha_{\mu}$'s are the eigenvalues of $A$ and the $c_{\mu}$'s
the fermionic operators associated to the corresponding eigenvectors. %
}. We assume here that both the Hamiltonian and the observable conserve
particle number. The more general case can be obtained by performing
a particle-hole transformation on some sites and considering a more
general covariance matrix. Exploiting the quadratic nature of the
problem and introducing the covariance matrix $R_{y,x}:=\tr\left(\rho_{0}c_{x}^{\dagger}c_{y}\right)$
($0\le R\le\1$) one can show that the expectation value $\mathsf{A}\left(t\right)$
reduces to a trace in the one-particle space: 
\begin{equation}
\mathsf{A}\left(t\right)=\tr\left(ae^{-itM}Re^{itM}\right)\,.\label{eq:A_free}
\end{equation}
Eq.~(\ref{eq:A_free}) is perfectly analogous to its many-body version
$\mathsf{A}\left(t\right)=\tr\left(Ae^{-itH}\rho_{0}e^{itH}\right)$
with $R$ playing the role of the initial state $\rho_{0}$. There
is however one importance difference: while $\tr\rho_{0}=1$ one has
$\tr R=N=\nu V$, i.e.~is \emph{extensive} (we defined $\nu=N/V$
the filling factor).

In the quasi-free setting the non-resonant condition necessary to
prove Eq.~(\ref{bound-varia}), does not hold. Let us then seek for
the analogous of the bound (\ref{bound-varia}) in this quasi-free
case. Let the one-particle Hamiltonian have the following diagonal
form $M=\sum_{k}\Lambda_{k}|k\rangle\langle k|$. The time averaged
covariance matrix is then $\overline{R}=\sum_{k}\langle k|R|k\rangle|k\rangle\langle k|$
(assuming non-degeneracy of the one-particle spectrum). We also define
$F_{k,q}=\langle k|a|q\rangle\langle q|R|k\rangle$. Assuming the
non-resonance condition for the one-particle spectrum, one gets $\Delta\mathsf{A}^{2}=\tr F^{2}-\sum_{k}\left(F_{k,k}\right)^{2}\le\tr F^{2}=\sum_{k,q}\left|\langle k|a|q\rangle\right|^{2}\left|\langle q|R|k\rangle\right|^{2}$.
Now $R$ is a non-negative operator and therefore it induces a (possibly
degenerate) scalar product which satisfies Cauchy-Schwarz inequality:
$\left|\langle q|R|k\rangle\right|^{2}=\left|\langle q|k\rangle_{R}\right|^{2}\le\langle q|q\rangle_{R}\langle k|k\rangle_{R}=\langle q|R|q\rangle\langle k|R|k\rangle$.
This leads us to

\begin{equation}
\Delta\mathsf{A}^{2}\le\tr\left(a\overline{R}a\overline{R}\right)\le\left\Vert a\right\Vert ^{2}\tr\overline{R}^{2}.\label{eq:bound_variance}
\end{equation}
Now, since $0\le\overline{R}\le\1$, $\tr\overline{R}^{2}\le\tr\overline{R}=\tr R=N$,
we finally obtain $\Delta\mathsf{A}^{2}\le\left\Vert a\right\Vert ^{2}\nu V\,.$
This result seems to hint at the fact that fluctuations in the quasi-free
setting are proportional to the system volume and are hence much larger
than in the non-free case where they are exponentially small in $V$.
For an extensive operator $A$ one has $\overline{\mathsf{A}}=O(V)$
and so the bound (\ref{eq:bound_variance}) translates into $\Delta\mathsf{A}/\overline{\mathsf{A}}\le O(V^{-1/2})$.
However Eq.~(\ref{eq:bound_variance}) is just a bound and nothing
prevents, in principle, to have much smaller fluctuations. In the
following we will provide arguments justifying, in this quasi-free
setting, extensivity for all the (temporal) cumulants.

Let us write again the generic expectation value (\ref{eq:A_free})
in the basis which diagonalizes $M$: 
\begin{equation}
\mathsf{A}\left(t\right)=\overline{\mathsf{A}}+2\sum_{k<q}\left|F_{k,q}\right|\cos\left(t\left(\Lambda_{k}-\Lambda_{q}\right)+\phi_{k,q}\right)
\end{equation}
 with $\phi_{k,q}=\arg F_{k,q}$. Consider the moment generating function
of $\mathsf{A}-\overline{\mathsf{A}}$ $\chi_{A}\left(\lambda\right):=\overline{e^{\lambda\left(\mathsf{A}\left(t\right)-\overline{\mathsf{A}}\right)}}$.
The derivatives of $\chi_{A}$ at $\lambda=0$ are precisely the (centered)
temporal moments of the random variable $\mathsf{A}(t)$. Now we observe
that \emph{if the (one-particle) energies are rationally independent}
(i.e.~linearly independent on the field of rationals), as a consequence
of the theorem of the averages, infinite time averages are the same
as uniform averages over the torus $\mathbb{T}^{V}$. In particular
one has

\begin{equation}
\overline{e^{\lambda\left(\mathsf{A}\left(t\right)-\overline{\mathsf{A}}\right)}}=\left(\prod_{j=1}^{V}\int\frac{d\theta_{j}}{2\pi}\right)\exp{\left[\lambda E\left(\boldsymbol{\theta}\right)\right]}.
\end{equation}

The generating function $\chi_{A}\left(\lambda\right)$ is exactly
given by the partition function of the generalized, classical XY model
with energy $E\left(\boldsymbol{\theta}\right)=2\sum_{k<q}\left|F_{k,q}\right|\cos\left(\theta_{k}-\theta_{q}+\phi_{k,q}\right)$
and inverse temperature $\beta=-\lambda$. The matrix $\left|F_{k,q}\right|$
defines the lattice of the interactions while the phases $\phi_{k,q}$
give the offset from which the angles are measured. Note that the
behavior of the density $P_{A}\left(a\right)$ is dictated by $\chi_{A}\left(\lambda\right)$
in a neighborhood of $\lambda=0$ which corresponds to infinite temperature
of the classical XY model.

\begin{figure}
\begin{centering}
\includegraphics[width=5cm,height=35mm]{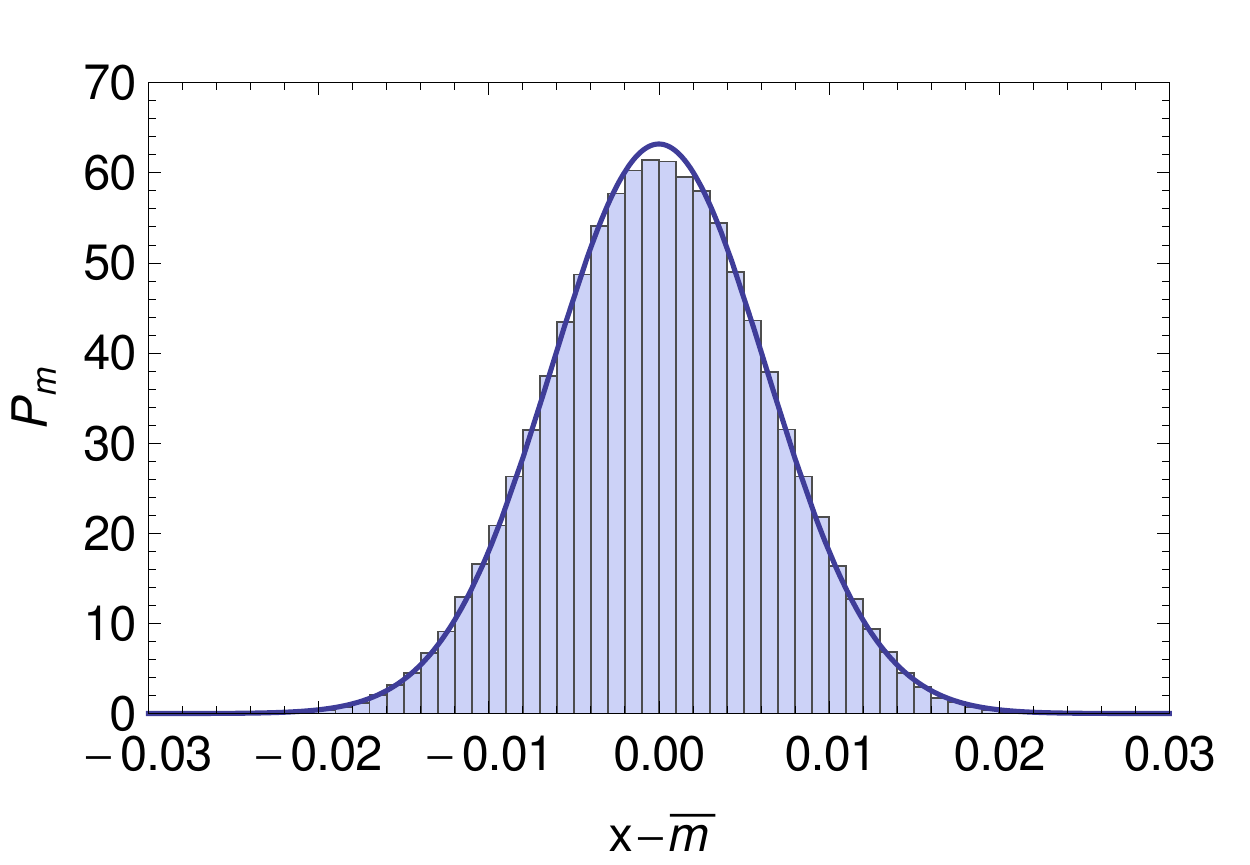} \includegraphics[width=5cm,height=35mm]{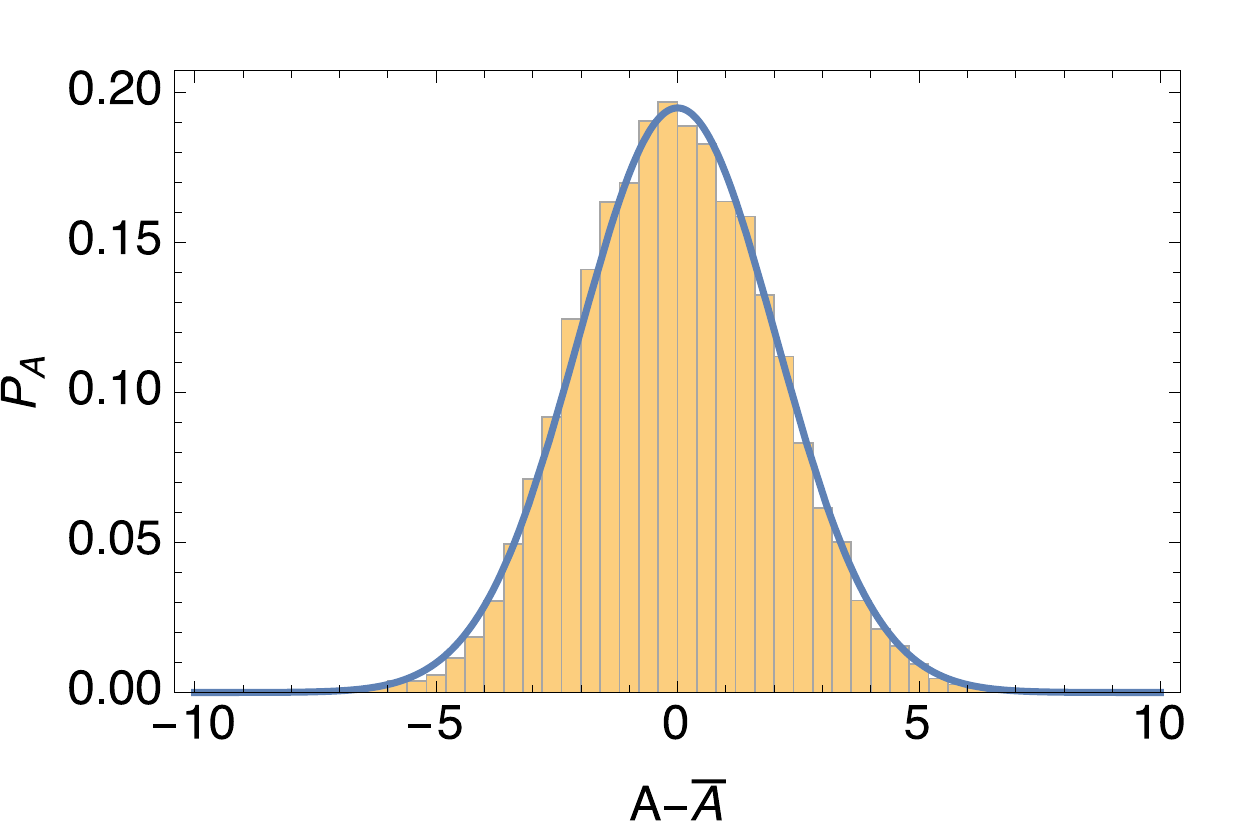} 
\par\end{centering}

\protect\protect\caption{Observation of Gaussian equilibration in quasi-free systems. Left
panel: Ising model in transverse field. The observable is the transverse
magnetization per site $m\left(t\right)=\langle\sigma_{i}^{z}\left(t\right)\rangle$.
Right panel: tight-binding model withe twisted periodic boundary conditions.
The observable is $A=\sum_{x=1}^{\ell}c_{x}^{\dagger}c_{x}$ where
$\ell<L$ is only a fraction of the total system size $L$. The thick
curves are the theoretical Gaussian prediction. Taken from Ref.~\protect\refcite{campos_venuti_gaussian_2013}.
\label{fig:Gaussian_equilibration}}
\end{figure}

A similar mapping can also be obtained in the general (i.e.~non quasi-free)
case. However the one-particle space has a natural underlying geometric
structure. For instance, the labels $k,q$ represent points in momentum
(real) space in a superfluid (localized) phase and the distance $\left|k-q\right|$
is well defined. Now, when\emph{ }the matrix elements $\left|F_{k,q}\right|$
decay sufficiently fast as $\left|k-q\right|\to\infty$ the corresponding
XY model is well defined in the thermodynamic limit, i.e.~the intensive
free energy has a limit as $L\to\infty$. This happens for instance
in case $\left|F_{k,q}\right|$ decays exponentially in $\left|k-q\right|$
or if one has $\left|F_{k,q}\right|\sim1/\left|k-q\right|^{\gamma}$
with $\gamma>D$. When this is the case one has $\chi_{A}\left(\lambda\right)=\exp L^{D}f\left(\lambda\right)$
where $f\left(\lambda\right)$ is the free energy per site. Moreover,
under these conditions, one expects $f\left(\lambda\right)$ to be
analytic in the high temperature, $\lambda=0$, limit, implying that
all the cumulants of $\mathsf{A}\left(t\right)$ are extensive. From
this we immediately draw the central limit theorem: as $L\to\infty$
the variable $(\mathsf{A}\left(t\right)-\overline{\mathsf{A}})/L^{D/2}$
tends in distribution to a Gaussian with zero mean and finite variance
given by $\partial_{\lambda=0}^{2}f\left(\lambda\right)/2$. This
situation has been referred to as \emph{Gaussian equilibration} in
Ref.~\refcite{campos_venuti_gaussian_2013}. We would like to stress
here that the term Gaussian equilibration refers to the situation
where all the temporal cumulants of an extensive observable scale
as the system's volume. In this sense one cannot have Gaussian equilibration
in the non-free setting since in that case the variance is exponentially
small in the volume. However it is still possible that the properly
rescaled variable $(\mathsf{A}\left(t\right)-\overline{\mathsf{A}})/\Delta\mathsf{A}$
, converges to a Gaussian in the infinite volume limit also for generic
truly interacting systems. 

The above arguments can be more precise for specific models and even
transformed into theorems (see Ref.~\refcite{campos_venuti_gaussian_2013}).
A manifestation of Gaussian equilibration is shown in Fig.~\ref{fig:Gaussian_equilibration}

\begin{figure}
\noindent \begin{centering}
\includegraphics[width=78mm,height=55mm]{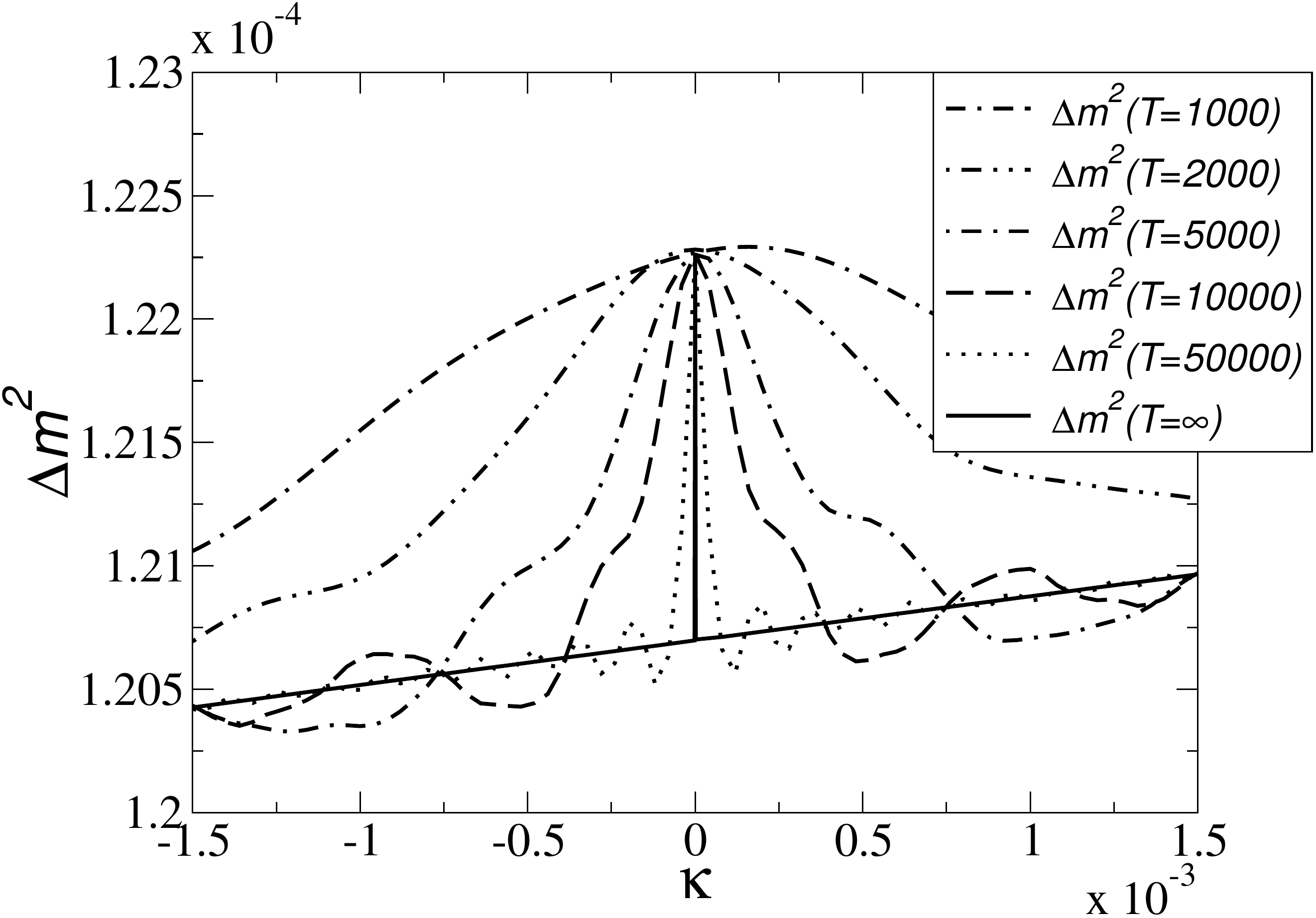} 
\par\end{centering}

\protect\protect\caption{Variance of $m\left(t\right)=\langle\sigma_{i}^{z}\left(t\right)\rangle$
for the model (\ref{eq:TAM}), as a function of the integrability
breaking parameter $\kappa$. The quench parameters are $(\kappa_{0}=0,h_{0}=2.0),\,\to(\kappa_{1}=\kappa,\, h_{1}=2.7)$.
The size is $L=8$. Dashed curves refers to the time variance computed
with a finite observation window $[0,T]$. \label{fig:var_kappa} }

\vspace{6mm}
 
\end{figure}

The above discussion shows that temporal fluctuations, and in particular
the variance, can be used to detect proximity to an integrable point.
Imagine a model which becomes integrable when an external parameter
$\kappa$ becomes, say, zero. Since temporal variance is expected
to be exponentially larger at an integrable point, it must be a discontinuous
function of $\kappa$ at the integrable point. This predictions are
confirmed by numerical simulations on the following model

\begin{equation}
H=-\sum_{i=1}^{L}\left[\sigma_{i}^{x}\sigma_{i+1}^{x}+h\sigma_{i}^{z}-\kappa\sigma_{i}^{x}\sigma_{i+2}^{x}\right]\label{eq:TAM}
\end{equation}
which is non-integrable for all values $\kappa\neq0$ (see Fig.~\ref{fig:var_kappa}).

\section{Temporal fluctuations after a small quench\label{sub:Temporal-fluctuations-after}}

Another situation where the temporal fluctuations can be characterized
in some generality is that of a small quench experiment. The system
is prepared in the ground state of the Hamiltonian $H_{0}$ for $t<0$.
At time $t=0$ one suddenly switches on a small perturbation $B$
such that the evolution Hamiltonian becomes $H=H_{0}+\delta\lambda B$,
with $\delta\lambda$ a small parameter. The small quench condition
can be found requiring that the exponentially small bound on the variance
does not hold, i.e.~$\tr(\overline{\rho}^{2})\nsim e^{-\alpha V}$.
For small quench $\tr(\overline{\rho}^{2})$ can be related to the
fidelity susceptibility $F$, more precisely one has $\tr(\overline{\rho}^{2})\simeq F^{4}$
\cite{rossini_decoherence_2007}. The scaling behavior of the fidelity
has been predicted in Ref.~\refcite{campos_venuti_quantum_2007}. At
regular point of the phase diagram, $F\sim\exp[-\alpha'\delta\lambda^{2}V],$
whereas in the critical region the scaling becomes $F\sim\exp[-\alpha''\delta\lambda^{2}L^{2/\nu}]$,
where $\nu$ is the correlation length critical exponent (defined
by $\xi\sim|\lambda-\lambda_{c}|^{-\nu}$). Summarizing, the small
quench condition reads $\delta\lambda\ll L^{-D/2}$ at regular points,
or $\delta\lambda\ll L^{-1/\nu}$ if $H_{0}$ is at a critical point.

\paragraph*{Finite size}

\begin{figure}
\noindent \begin{centering}
\includegraphics[width=4.2cm,height=2.5cm]{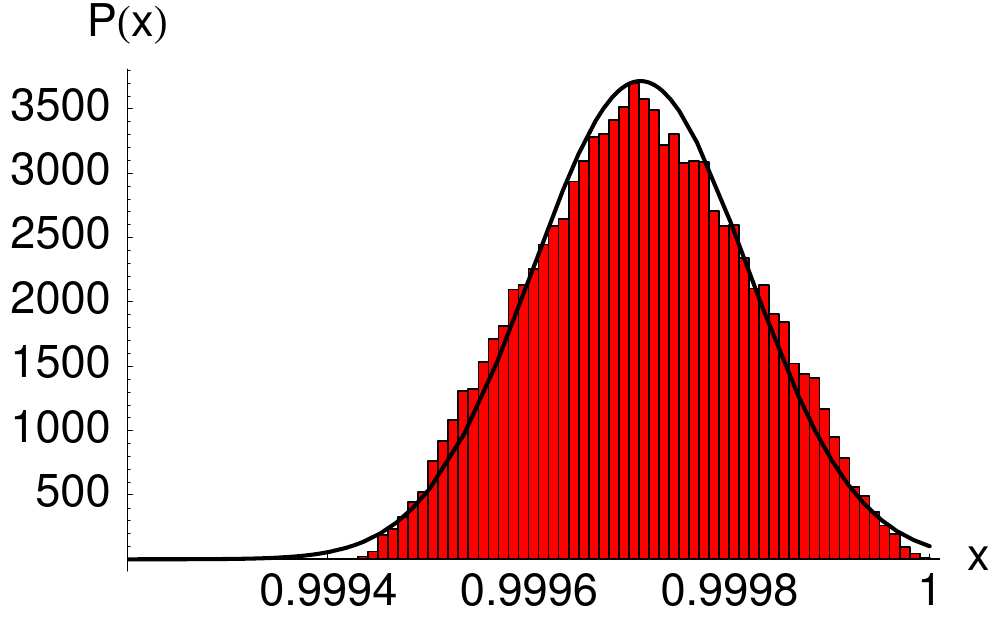}
\includegraphics[width=4.2cm,height=2.5cm]{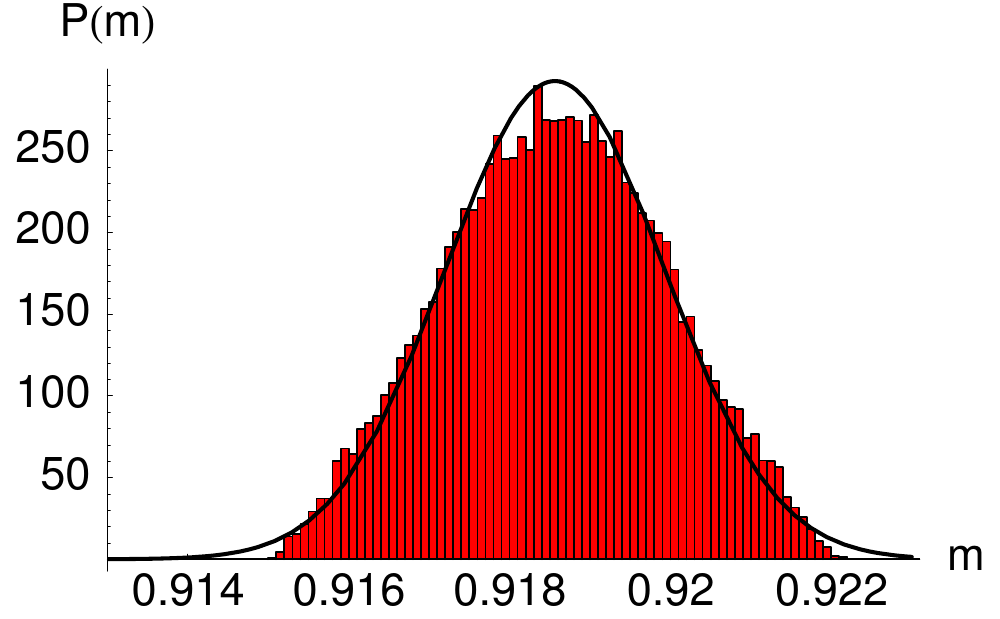} 
\par\end{centering}

\noindent \begin{centering}
\includegraphics[width=4.2cm,height=2.5cm]{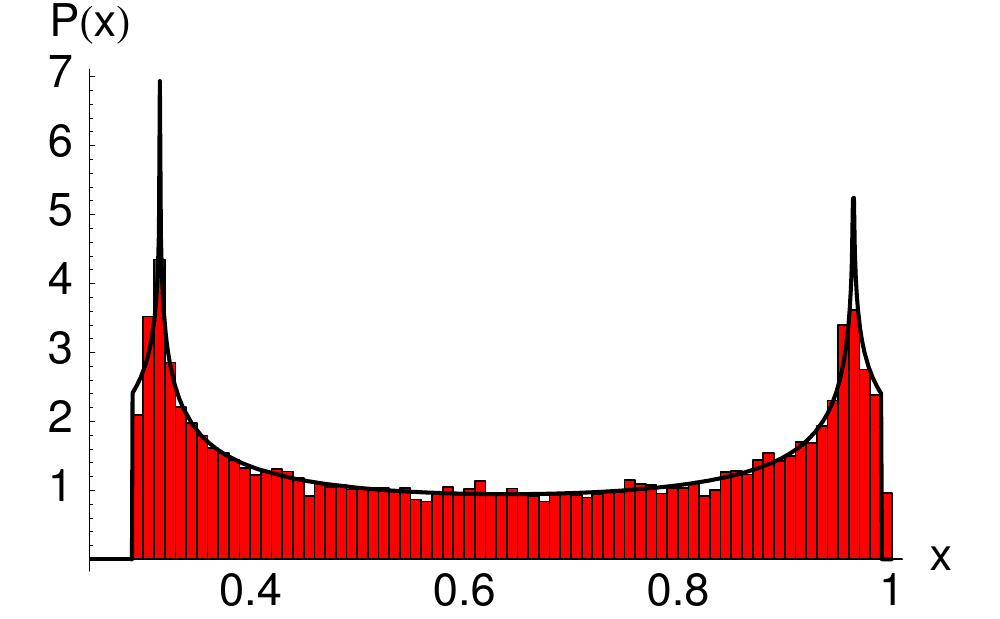} \includegraphics[width=4.2cm,height=2.5cm]{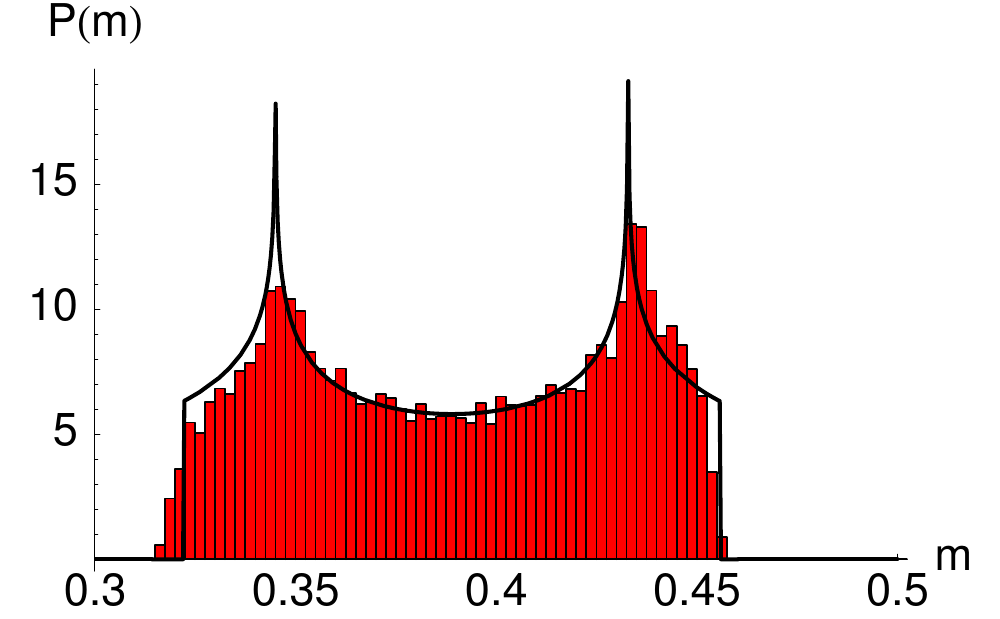} 
\par\end{centering}

\protect\protect\caption{{\small{}{}
Probability distributions for a small quench. $P\left(x\right)=\overline{\delta\left(x-\mathcal{L}\left(t\right)\right)}$
and $P\left(m\right)=\overline{\delta\left(m-\langle\sigma_{1}^{z}\left(t\right)\rangle\right)}$
refer to the Loschmidt echo (left panels) and magnetization respectively
(right panels). Upper panels: the quench is performed at a regular
point of the phase diagram. Lower panels: the same quench amplitude
performed close to a quantum critical point. Note the much larger
scale of the horizontal axis. The thick lines are our analytic predictions
using only the three largest weights. The Hamiltonian is a non-integrable
extension of the Ising model in transverse field. Sizes are $L=12(16)$
for the upper (lower) panel. See \protect\cite{campos_venuti_universality_2010,campos_venuti_exact_2011} for details.} 
\label{fig:batman}}
\end{figure}

Let us now consider the full temporal statistic of $\mathsf{A}(t)$
at a fixed size and fixed --small-- quench. The result for a generic
observable is shown in Fig.~\ref{fig:batman}. In the off-critical
region, $\xi_{i/f}\ll L$ ($\xi_{i/f}$ denotes the correlation length
of the initial/final Hamiltonian), full temporal statistics of generic
observables are approximately Gaussian. In the complementary, quasi-critical
region $\xi_{i/f}\gg L$, the distribution becomes bimodal characterized
by larger variance. This result is completely general, it holds both
for integrable and non-integrable models and generic observables\cite{campos_venuti_unitary_2010,campos_venuti_universality_2010}.
As can be seen from Fig.~\ref{fig:batman}, this effect is significant
even for small system sizes of the order of $L\sim10$, and can be
used to locate quantum critical points using out-of-equilibrium methods.
This is an important point as it is notoriously difficult to observe
precursor of quantum phase transitions with such short sizes using
equilibrium indicators. The explanation of this phenomenon is the
following. Consider the general form of an observable expectation
value 
\begin{equation}
\mathsf{A}(t)=\sum_{n,m}e^{-it(E_{n}-E_{m})}A_{m,n}\langle n|\psi_{0}\rangle\langle\psi_{0}|m\rangle.
\end{equation}
Noting that, $\langle\psi_{0}|m\rangle=O(\delta\lambda)$ for $m>0$
and $\langle\psi_{0}|0\rangle=1+O(\delta\lambda^{2})$, in the small
quench regime one has approximately 
\begin{equation}
\mathsf{A}(t)\simeq\overline{\mathsf{A}}+\Big[\sum_{n>0}e^{-it(E_{n}-E_{0})}A_{0,n}\langle n|\psi_{0}\rangle+\mathrm{c.c.}\Big].\label{eq:approx1}
\end{equation}
If the energy gaps $(E_{n}-E_{0})$ are rationally independent one
can show that the random variable $\mathsf{A}(t)$ in Eq.~(\ref{eq:approx1})
is a sum of independent random variables: $\mathsf{A}(t)-\overline{\mathsf{A}}=\sum_{n>0}X_{n}$.
Each $X_{n}$ has probability distribution function given by $1/(\pi\sqrt{\sigma_{n}^{2}-x^{2}})$
with variance $\sigma_{n}^{2}=2\left|A_{0,n}\langle n|\psi_{0}\rangle\right|^{2}$.
If $H_{0}$ is at a regular point of the phase diagram, the coefficients
$\langle n|\psi_{0}\rangle$ (and $A_{0,n}$) have no particular structure.
Correspondingly one expects that none of them dominate and essentially
$\mathsf{A}(t)$ will be Gaussian distributed. Note that, in this
regime the total variance satisfies 
\begin{equation}
\Delta\mathsf{A}^{2}=\sum_{n>0}\sigma_{n}^{2}\le2\sum_{n>0}A_{0,n}A_{n,0}=2[\langle A^{2}\rangle-\langle A\rangle^{2}].
\end{equation}
The last term is precisely (twice) the quantum variance computed with
state $|\psi_{0}\rangle$ (or $|0\rangle$ as they give the same result
up to $O(\delta\lambda^{2})$). As such, for an extensive observable,
the temporal variance is bounded by the volume $\Delta\mathsf{A}^{2}\le O(V)$
in this regime.

The situation is different if $H_{0}$ (or $H$) is close to a quantum
critical point. In this case one can show that $\langle E|\psi_{0}\rangle\sim\delta\lambda E^{-1/(\zeta\nu)}$
($\zeta$ is the dynamical critical exponent) \cite{campos_venuti_universality_2010,de_grandi_quench_2010}.
At finite size, the lowest modes have energy, $E_{n}=v\left(2\pi n/L\right)^{\zeta}$
so that $\langle E_{n}|\psi_{0}\rangle\sim\delta\lambda L^{1/\nu}$.
In practice, since in the region of validity of perturbation theory,
$\langle E_{0}|\psi_{0}\rangle$ is already ``large'', the sum rule
$\sum_{n}\left|\langle E_{n}|\psi_{0}\rangle\right|^{2}=1$ constrains
to have only very few $\langle E_{n}|\psi_{0}\rangle$ appreciably
different from zero. In practice, a good approximation is obtained
retaining only the two dominant terms in the sum in Eq.~(\ref{eq:approx1}).
Assuming for simplicity that $A_{0,n}\langle n|\psi_{0}\rangle$ are
real, one has approximately, in the quasi-critical regime 
\begin{equation}
\mathsf{A}(t)\simeq\overline{\mathsf{A}}+W_{1}\cos(t\omega_{1})+W_{2}\cos(t\omega_{2}),\label{eq:A_critic}
\end{equation}
with $W_{n}=2A_{0,n}\langle n|\psi_{0}\rangle$. The (temporal) probability
density corresponding to Eq.~(\ref{eq:A_critic}) is precisely the
density of states of a two dimensional, anisotropic, tight-binding
model with hopping constants $W_{1},\, W_{2}$. It has been computed
analytically in Ref.~\refcite{campos_venuti_unitary_2010}. The resulting
distribution $P_{A}(a)$ is symmetric around the mean $\overline{\mathsf{A}}$,
supported in $\left[\overline{\mathsf{A}}-\left|\left|W_{1}\right|+\left|W_{2}\right|\right|,\overline{\mathsf{A}}+\left|\left|W_{1}\right|+\left|W_{2}\right|\right|\right]$
with logarithmic divergences at $a=\overline{\mathsf{A}}\pm\left|\left|W_{1}\right|-\left|W_{2}\right|\right|$
(see Fig.~\ref{fig:batman} lower panels).

\paragraph*{Thermodynamic limit}

If one keeps the quench strength $\delta\lambda$ fixed, and increases
the system size $L$, one will eventually enter the off-critical region.
Increasing $L$ further one will leave the perturbative, small-quench,
region and temporal variances will become exponentially small in the
system size. Nevertheless one can wonder whether it is possible to
obtain a meaningful limit, by keeping $\delta\lambda$ in the small
quench region, and sending $L$ to infinity. This situation has been
studied in Ref.~\refcite{campos_venuti_universal_2014}.

Expanding $\mathsf{A}(t)$ up to first order in $\delta\lambda$ using
Dyson expansion and the spectral resolution $H_{0}=\sum_{n}E_{n}|n\rangle\langle n|$,
one gets 
\begin{equation}
\mathsf{A}(t)=\overline{\mathsf{A}}+\delta\lambda\sum_{n>0}\left(Z_{n}e^{-it(E_{n}-E_{0})}+\mathrm{c.c.}\right)+O\left(\delta\lambda^{2}\right),\label{eq:full_expansion}
\end{equation}
where the first, time-independent term is the average of $\mathsf{A}\left(t\right)$
and with $Z_{n}:=A_{0,n}B_{n,0}/\left(E_{n}-E_{0}\right)$ and the
notation $A_{n,m}=\langle n|A|m\rangle$. The leading contribution
to the temporal variance is therefore at second order and assuming
that the gaps $E_{n}-E_{0}$ are non-degenerate one obtains 
\begin{equation}
\Delta\mathsf{A}_{B}^{2}=2\delta\lambda^{2}\sum_{n>0}\left|Z_{n}\right|^{2}+O\left(\delta\lambda^{3}\right).\label{eq:var_AB}
\end{equation}
We added a subscript $B$ to recall that the variance is computed
with perturbation $B$.

Using Eq.~(\ref{eq:full_expansion}) we can actually obtain the full
probability distribution of the variable $\mathsf{A}$. Assuming rational
independence of the gaps $E_{n}-E_{0}$ and using the theorem of averages
we obtain the following expression for the characteristic function
of $\mathsf{A}$, 
\begin{equation}
\overline{e^{is(\mathsf{A}-\overline{\mathsf{A}})/\delta\lambda}}=\prod_{n>0}J_{0}\left(2s\left|Z_{n}\right|\right):=J_{A}\left(s\right),\label{eq:chi_obs}
\end{equation}
where $J_{0}$ is the Bessel function of the first kind. So the probability
distribution of $\mathsf{A}$ is completely encoded in the characteristic
function $J_{A}\left(s\right)$. The cumulants of the variable $(\mathsf{A}-\overline{\mathsf{A}})/\delta\lambda$
are given by $\kappa_{2p}=a_{2p}2^{2p}Q_{2p}$ with $Q_{2p}:=\sum_{n>0}\left|Z_{n}\right|^{2p}$
and known constants $a_{p}$ %
\footnote{The coefficients $a_{p}$ are defined by the series $\ln[I_{0}(s)]=\sum_{p=1}^{\infty}a_{p}s^{p}/n!$
which converges absolutely in a neighborhood of the origin. $I_{0}$
is a modified Bessel function. Note that $a_{p}=0$ for $p$ odd. %
} (odd cumulants are zero). Under the assumption of convergence the
probability distribution of $\mathsf{A}$ is uniquely characterized
by the coefficients $Q_{2p}$. Conversely the probability distribution
uniquely defines the coefficients $Q_{2p}$ which are generalizations
of the variance Eq.~(\ref{eq:var_AB}). Intuitively, at critical
points the cumulants $\kappa_{2p}$ (through the coefficients $Q_{2p}$)
may diverge with the system size. 

Let us analyze the behavior of $Q_{2p}$ close to quantum criticality.
In this case $\delta\lambda=\left|\lambda-\lambda_{c}\right|$ measures
the distance from the critical point $\lambda_{c}$. Using standard
scaling arguments one can show that $Q_{2p}\propto L^{2p\alpha}$
with $\alpha=2D+\zeta-\Delta_{A}-\Delta_{B}$ (see Ref.~\refcite{campos_venuti_universal_2014}
for details). Here $\Delta_{A/B}$ are the scaling dimensions of the
observables $A/B$ that we assumed extensive. Instead, away from criticality
the expectation is $Q_{2p}\propto L^{D}$. Requiring that, at finite
size, $Q_{2p}$ is analytic in the system parameters and matches the
above scaling, one can predict the behavior of $Q_{2p}$ close to
the critical point both in the critical region $\xi\gg L$ and in
the off-critical one $\xi\ll L$ %
\footnote{The notation $f(L)\sim g(L)$ means that $\lim_{L\to\infty}f(L)/g(L)=M$
for some constant $M$.%
}: 
\begin{equation}
\kappa_{2p}\propto Q_{2p}\sim\begin{cases}
L^{2\alpha p} & \xi\gg L\\
\delta\lambda^{D\nu-2\alpha p\nu}L^{D} & \xi\ll L
\end{cases}\,.\label{eq:Q_scaling}
\end{equation}
As usual in finite size scaling theory, the above prediction refers
to the singular part of $Q_{2p}$, on top of which there is always
a regular, extensive, contribution \cite{domb_phase_1983}.

\begin{figure}
\begin{centering}
\includegraphics[width=7cm,height=5cm]{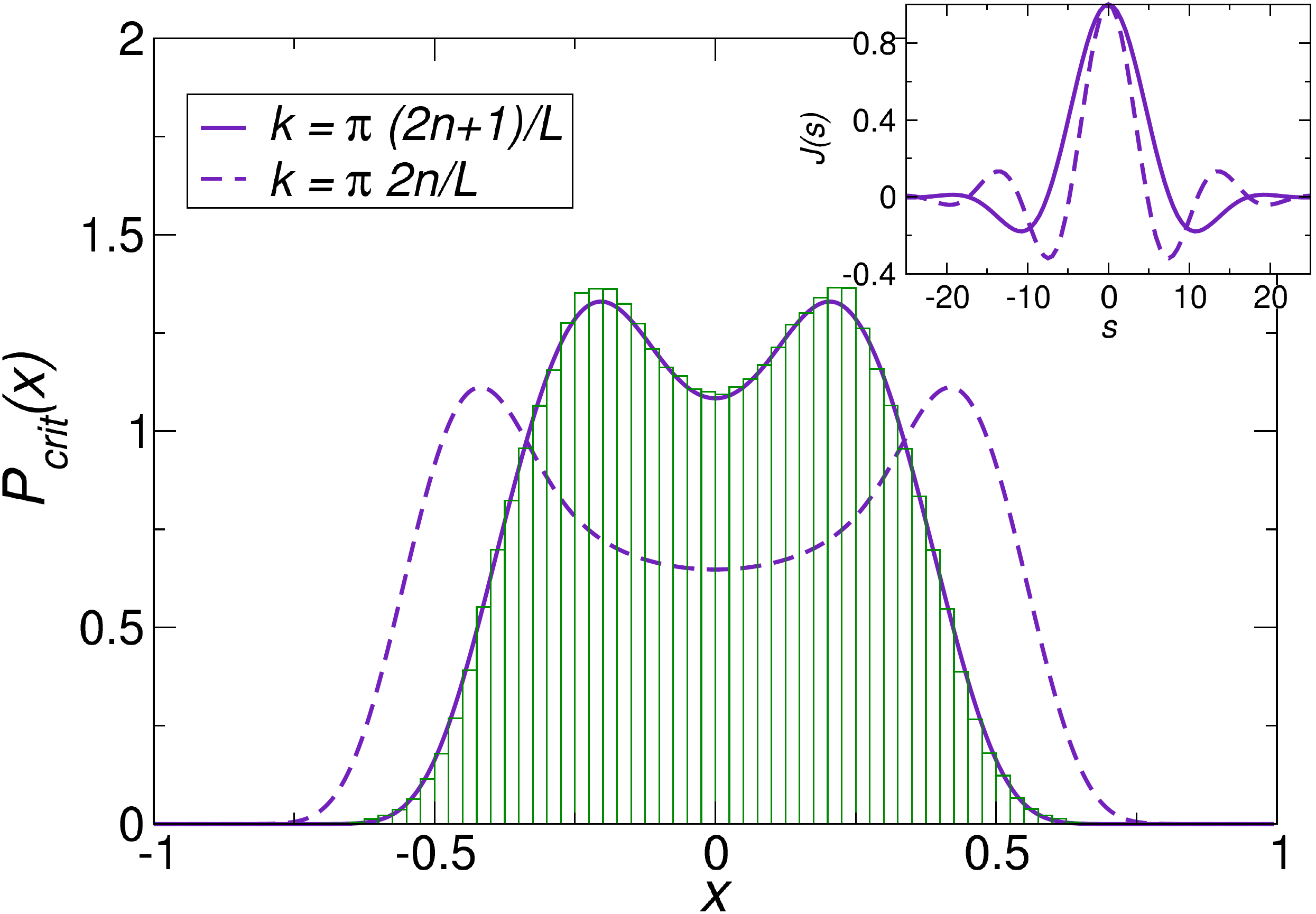} 
\par\end{centering}

\protect\protect\caption{Critical probability distribution for the transverse magnetization
$[\langle M(t)\rangle-\overline{M}]/(L\delta h)$ in the XY-model
in transverse field. The inset shows the characteristic function,
which in the thermodynamics limit becomes $J\left(s\right)=\prod_{n=0}^{\infty}J_{0}\left(s/\left(2n+1\right)\right)$.
The histogram is computed performing a numerical experiment on the
Ising model on a chain of $L=1006$ sites with periodic boundary conditions
corresponding to quasimomenta of the form $k=\pi(2n+1)/L$. The continuous
(dashed) lines refer to quasimomenta of the form $k=\pi(2n+1)/L$
($k=2\pi n/L$). The quench parameters are $h_{1}=1$, $h_{2}=1.0003$
and $\gamma_{1}=\gamma_{2}=1$. The statistics is obtained sampling
600,000 random times uniformly distributed in $\left[0,T\right]$
with $T=600,000$. The distribution is unchanged using a different
$\gamma_{1}=\gamma_{2}\protect\neq0$ as implied by universality.
From \protect\cite{campos_venuti_universal_2014}.
\label{fig:FTS-critical}}
\end{figure}

Consider now the rescaled random variable $\mathsf{X}\left(t\right)=(\mathsf{A}\left(t\right)-\overline{\mathsf{A}})/\Delta\mathsf{A}$
whose cumulants are given by $\kappa_{2n}^{X}=\kappa_{2n}^{A}/(\kappa_{2}^{A})^{n}$
for $n\ge1$ whereas odd cumulants are zero. The probability distribution
of $\mathsf{X}$ is uniquely determined by the ratios $R_{2p}=Q_{2p}/\left(Q_{2}\right)^{p}$.
From Eq.~(\ref{eq:Q_scaling}) we see that in the quasi-critical
regime, these ratios are scale independent and define some presumably
universal constants. Let us now find these constants. With the help
of density of states $\rho\left(E\right)=\tr\left(\delta\left(H-E\right)\right)$
we can write $Q_{p}=\int Q_{p}\left(E\right)\rho\left(E\right)dE$.
Since $\rho\left(E\right)dE$ is scale invariant, from $Q_{p}\propto L^{p\alpha}$
we derive $Q_{p}\left(E\right)\propto E^{-p\alpha/\zeta}$. We now
assume that at the critical point one has vanishing energy excitations
with definite momentum. In order to proceed further we must specify
the form of the low energy dispersion. The simplest possibility is
a rotationally invariant spectrum at small momentum, i.e.~$E\simeq C\left\Vert \boldsymbol{k}\right\Vert ^{\zeta}=C(\sum_{j}k_{j}^{2})^{\zeta/2}$
where $\boldsymbol{k}$ is a quasi-momentum vector. In one dimension
this is essentially the only possibility but for $d>1$ one can also
have anisotropic transitions where the form of the dispersion depends
on the direction. Using the isotropic assumption we obtain $Q_{p}\simeq C'\sum_{\boldsymbol{k}}\left\Vert \boldsymbol{k}\right\Vert ^{-p\alpha}.$
In doing so we have essentially restricted the sum over $n$ to the
one-particle contribution. This is expected to be the leading contribution
whereas higher particle sectors contribute at most to the extensive,
regular term \cite{vasseur_crossover_2013}. This shows that the
cumulants of $\mathsf{X}$ are \emph{uniquely} specified by the critical
exponent $\alpha$ and the \emph{boundary conditions} that specify
$\boldsymbol{k}$. More precisely the probability distribution of
the rescaled variable $\mathsf{X}\left(t\right)$ is a universal function
which depends only on $\alpha$ and the boundary conditions. A related
universal behavior has been observed in \cite{dalla_torre_universal_2013,mitra_correlation_2013}
in the case of the sine-Gordon model. Let us assume for concreteness
that the lattice is a hyper-cube of size $L$ and the boundary conditions
(BC) are such that moments are quantized according to $\boldsymbol{k}=(2\pi/L)(\boldsymbol{n}+\boldsymbol{b})$
with $n_{i}=1,\ldots,L$. The BC on the direction $i$ are fixed by
$b_{i}\in\left[0,1/2\right]$ which interpolates between periodic
(PBC, $b_{i}=0$) and anti-periodic (ABC, $b_{i}=1/2$) BC. In the
infinite volume limit the ratios $R_{2p}=Q_{2p}/\left(Q_{2}\right)^{p}$
become universal quantities that can be explicitly computed (see \cite{campos_venuti_universal_2014}).
The result is 
\begin{equation}
\lim_{L\to\infty}R_{2p}=\begin{cases}
\delta_{p,1} & 2\alpha\le D\\
\zeta_{\boldsymbol{b}}\left(2p\alpha\right)/\zeta_{\boldsymbol{b}}\left(2\alpha\right)^{p} & 2\alpha>D.
\end{cases}\label{eq:ratios}
\end{equation}
where $\zeta_{\boldsymbol{b}}\left(\alpha\right)=\sum_{n_{1}=1}^{\infty}\cdots\sum_{n_{d}=1}^{\infty}\left\Vert \boldsymbol{n}+\boldsymbol{b}\right\Vert ^{-\alpha}$
denotes a generalized $D$-dimensional Hurwitz-Epstein $\zeta$-function.
For $2\alpha\le D$ the characteristic function of $\mathsf{X}(t)$
becomes $e^{-s^{2}/2}$ in the thermodynamic limit and so $\mathsf{X}$
tends in distribution to Gaussian. Clearly the Gaussian behavior observed
here for not sufficiently relevant operators, i.e.~$\alpha\le D/2$,
is also to be expected at regular points of the phase diagram. These
predictions have been checked for the XY model in transverse field\cite{campos_venuti_universal_2014}.
Considering the transverse magnetization $\mathsf{M}\left(t\right)=\sum_{j}\langle\sigma_{j}^{z}(t)\rangle$
as observable, its temporal characteristic function can be computed
analytically. The scaling dimensions in this case are $D=\zeta=\Delta_{A}=\Delta_{B}=1$
implying $\alpha=1$. One can then prove analytically that in the
limit $L\to\infty$, Eq.~(\ref{eq:chi_obs}) becomes $\prod_{n=0}^{\infty}J_{0}(\lambda/\alpha_{n})$
with $\alpha_{n}=\sqrt{\zeta_{1/2}\left(2\right)/2}\,(n+1/2)$. This
in turns implies Eq.~(\ref{eq:ratios}) with $\alpha=1$ and $b=1/2$
as predicted. A numerical demonstration is provided in Fig.~\ref{fig:FTS-critical}.
A discussion of the regular points as well as a comparison of the
dynamical central limit type theorem here discussed and the one for
quantum fluctuations at equilibrium can be found in Ref.~\refcite{campos_venuti_universal_2014}.

\section{Temporal Fluctuations in non-homogeneous systems}

The results of the previous section, valid for homogeneous systems,
indicate that the temporal fluctuations can be used as indicator of
quantum criticality. A by-product of Eq.~(\ref{eq:Q_scaling}) is
that, for a small quench (in the sense of Section \ref{sub:Temporal-fluctuations-after}),
the temporal fluctuations diverge as $\Delta\mathsf{A}_{B}^{2}\sim L^{2\alpha}$
in the quasi-critical region $\xi\gg L$. The recipe to estimate critical
points is standard (see e.g.~\cite{domb_phase_1983,roncaglia_finding_2008}).
Since in the off-critical region $L\gg\xi$, $\Delta\mathsf{A}_{B}^{2}$
is extensive (for extensive observable $A$ and perturbation $B$),
finite size pseudo critical point $g^{\ast}(L)$ can be defined as
the location of the maximum of $\Delta\mathsf{A}_{B}^{2}$, as a function
of the tunable parameter $g$. The sequence $g^{\ast}(L)$ converges
to the exact critical point as $L\to\infty$. For this procedure to
succeed one must be able to identify a maximum, i.e.~one needs $2\alpha>D$.
We can now compare the ability of the the temporal variance to act
as indicator of quantum criticality with that of other standard, \emph{equilibrium,}
indicators such as quantum fluctuations $\Delta A^{2}:=\langle A^{2}\rangle-\langle A\rangle^{2}$,
or the generalized susceptibility $\chi_{AB}:=\left.\partial_{\lambda}\langle A\rangle_{\lambda}\right|_{\lambda=0}$,
(here $\langle\bullet\rangle_{\lambda}$ indicates the quantum average
taken with the ground state of $H(\lambda)=H_{0}-\lambda B$). Standard
scaling arguments allow to show that $\Delta A^{2}\sim L^{2d-2\Delta_{A}}$,
whereas $\chi_{AB}\sim L^{\alpha}$. Since all quantities are extensive
in the off-critical region, pseudo critical points can be defined
when $\Delta_{A}<D/2$ using quantum fluctuations, and for $\alpha>D$
in case of the susceptibility. Taking for simplicity $A=B$, pseudo-critical
points can be defined provided $\Delta_{A}<D/2$ (using quantum fluctuations),
$\Delta_{A}<D/2+\zeta/2$ (using generalized susceptibilities), $\Delta_{A}<(3/4)D+\zeta/2$
(using temporal variances). Since the smaller $\Delta_{A}$, the more
relevant is the operator $A$, we see that the above conditions are
less and less restrictive. In other words more quantum phase transitions
can be observed and located resorting to the temporal variance. 

The above considerations suggest that temporal variances may also
be used to detect phase boundaries between different phases in spatially
inhomogeneous systems. Suppose a large system can be divided in two
neighboring regions A and B, and that the system is in phase $P_{A}$
in region A and in phase $P_{B}$ in region B. For simplicity one
can think that the system is a very long one-dimensional chain, and
regions A, B, are two segments separated by a boundary region C. Since
the state of phase $P_{A}$ cannot be deformed continuously into the
state in B preserving the symmetries of the model, one expects that
some pseudo-critical behavior emerges in the boundary region separating
region A form region B. A particularly interesting example of such
inhomogeneous systems is provided by optical lattice experiments where
the presence of a (to a good approximation) harmonic confining potential
breaks translation invariance. Experiments \cite{gemelke_situ_2009,bakr_peng_10,sherson_weitenberg_10}
have been able to resolve the site-occupation profiles and reveal
the characteristic ``wedding cake'' structure in which Mott plateaus
are flanked by superfluid domains. For the purpose of accurately determining
the boundaries between those domains, several local, equilibrium,
compressibilities have been proposed in the literature \cite{batrouni_mott_2002,wessel_alet_04,rigol_local_2003,rigol_muramatsu_04},
including $\kappa_{i}:=\partial\langle\hat{n}_{i}\rangle/\partial\mu_{i}$
\cite{batrouni_mott_2002}, as well as the site-occupation fluctuations
$\Delta n_{i}^{2}:=\langle\hat{n}_{i}^{2}\rangle-\langle\hat{n}_{i}\rangle^{2}$
\cite{rigol_local_2003,rigol_muramatsu_04}, where $\langle\bullet\rangle$
stands for the quantum expectation value and $\mu_{i}$ is the local
chemical potential at site $i$. In Ref.~\refcite{yeshwanth_small_2014}
extensive numerical simulations confirmed that temporal fluctuations
of the site-occupation can serve as efficient detectors of this \emph{local
quantum criticality }\cite{rigol_local_2003}\emph{ }(see also \cite{natu_local_2011})\emph{,
}and in fact reveals details which cannot be observed using the equilibrium
local compressibility $\kappa_{i}$. We now briefly review those findings. 

We first consider a class of hard-core boson models, which can be
mapped to systems of interacting Fermions after Jordan-Wigner transformation.
This allows to perform numerical simulations on long chains and so
to obtain proper scaling of quantities with the system size $L$.
The Hamiltonian is given by
\begin{equation}
\hat{H}_{0}=-J\sum_{i=1}^{L-1}(\hat{b}_{i}^{\dagger}\hat{b}_{i+1}+\hc)+\sum_{i=1}^{L}[\lambda g_{i}\hat{n}_{i}+V_{0}(-1)^{i}\hat{n}_{i}].\label{eq:H_stag}
\end{equation}
which can be thought of as the limit $U/J\rightarrow\infty$ of the
Bose-Hubbard model \cite{cazalilla_citro_review_11}. In Eq.~\eqref{eq:H_stag},
$\hat{b}_{i}^{\dagger}$ ($\hat{b}_{i}$) is the creation (annihilation)
operator of a hard-core boson at site $i$, $\hat{n}_{i}=\hat{b}_{i}^{\dagger}\hat{b}_{i}$,
and $g_{i}$ describes a harmonic confining potential, with $g_{i}=L^{-2}(i-L/2+\epsilon)^{2}$.
The trap is shifted off-center by a small amount $\epsilon$ to remove
degeneracies in the energy levels and gaps of the Hamiltonian {[}see
the discussion of Eq.~\eqref{eq:qfree-var}{]}. We initialize the
system in a ground state $|\Psi(0)\rangle$ of a lattice with $L$
sites and $N$ hard-core bosons. After performing a sudden quench
on the trap potential, $\lambda\rightarrow\lambda+\delta\lambda$
at time $t=0$, the system evolves unitarily as $|\Psi(t)\rangle=\exp(-i\hat{H}t)|\Psi(0)\rangle$.
The post-quench Hamiltonian is given by $\hat{H}=\hat{H}_{0}+\delta\lambda\,\hat{B}$.
A Jordan-Wigner transformation maps Eq.~\eqref{eq:H_stag} onto a
Hamiltonian quadratic in fermion operators $\hat{f}_{i}^{\dagger}$
and $\hat{f}_{i}$. From that transformation, it follows that the
site occupations of hard-core bosons and spinless fermions are identical,
i.e.~$\hat{n}_{i}=\hat{b}_{i}^{\dagger}\hat{b}_{i}=\hat{f}_{i}^{\dagger}\hat{f}_{i}$.
The fermionic Hamiltonian can be written as $\hat{H}=\sum_{i,j}\hat{f}_{i}^{\dagger}M_{i,j}\hat{f}_{j}$
with $M_{i,j}=-J(\delta_{i,j+1}+\delta_{i,j-1})+\left[(\lambda+\delta\lambda)g_{i}+V_{0}(-1)^{i}\right]\delta_{i,j}$.
The noninteracting character of the fermionic system allows one to
write temporal fluctuations of site occupations (and in fact of any
quadratic observable in the fermions) in terms of one-particle quantities
alone. Consider a general quadratic observable of the form $\hat{X}=\sum_{i,j}\hat{f}_{i}^{\dagger}\Gamma_{i,j}\hat{f}_{j}$.
One can show that $\mathsf{X}(t)=\langle\Psi(t)|\hat{X}|\Psi(t)\rangle=\tr(\hat{X}e^{-it\hat{H}'}\rho_{0}e^{it\hat{H}'})=\tr(\Gamma e^{-itM}Re^{itM})$
where $R$ is the covariance matrix of the initial state $\rho_{0},$
i.e.,~$R_{i,j}=\tr(\rho_{0}\hat{f}_{j}^{\dagger}\hat{f}_{i})$ (note
that the initial state does not necessarily need to be Gaussian).
Let the one-particle Hamiltonian $M$ have the spectral representation
$M=\sum_{k}\Lambda_{k}|k\rangle\langle k|$ ($|k\rangle$ are one
the particle eigenfunctions) and define $F_{k,q}=\langle k|\Gamma|q\rangle\langle q|R|k\rangle$
where $\Gamma,\, R$ are one-particle operators. The temporal variance
of $\mathsf{X}$ is given by 
\begin{equation}
\Delta\mathsf{X}^{2}=\tr(F^{2})-\sum_{k}(F_{k,k})^{2}.\label{eq:qfree-var}
\end{equation}
 Equation~(\ref{eq:qfree-var}) relies on the assumption of the non-resonant
conditions for the one-particle spectrum \cite{reimann_foundation_2008,campos_venuti_gaussian_2013},
which has been verified in our numerical calculations (for $\epsilon\neq0$).
To compute the variance of the site occupations we simply take $X=n_{i}$
which implies $\Gamma_{x,y}^{(i)}=\delta_{i,x}\delta_{i,y}$.

\begin{figure}
\begin{centering}
\includegraphics[clip,width=8cm]{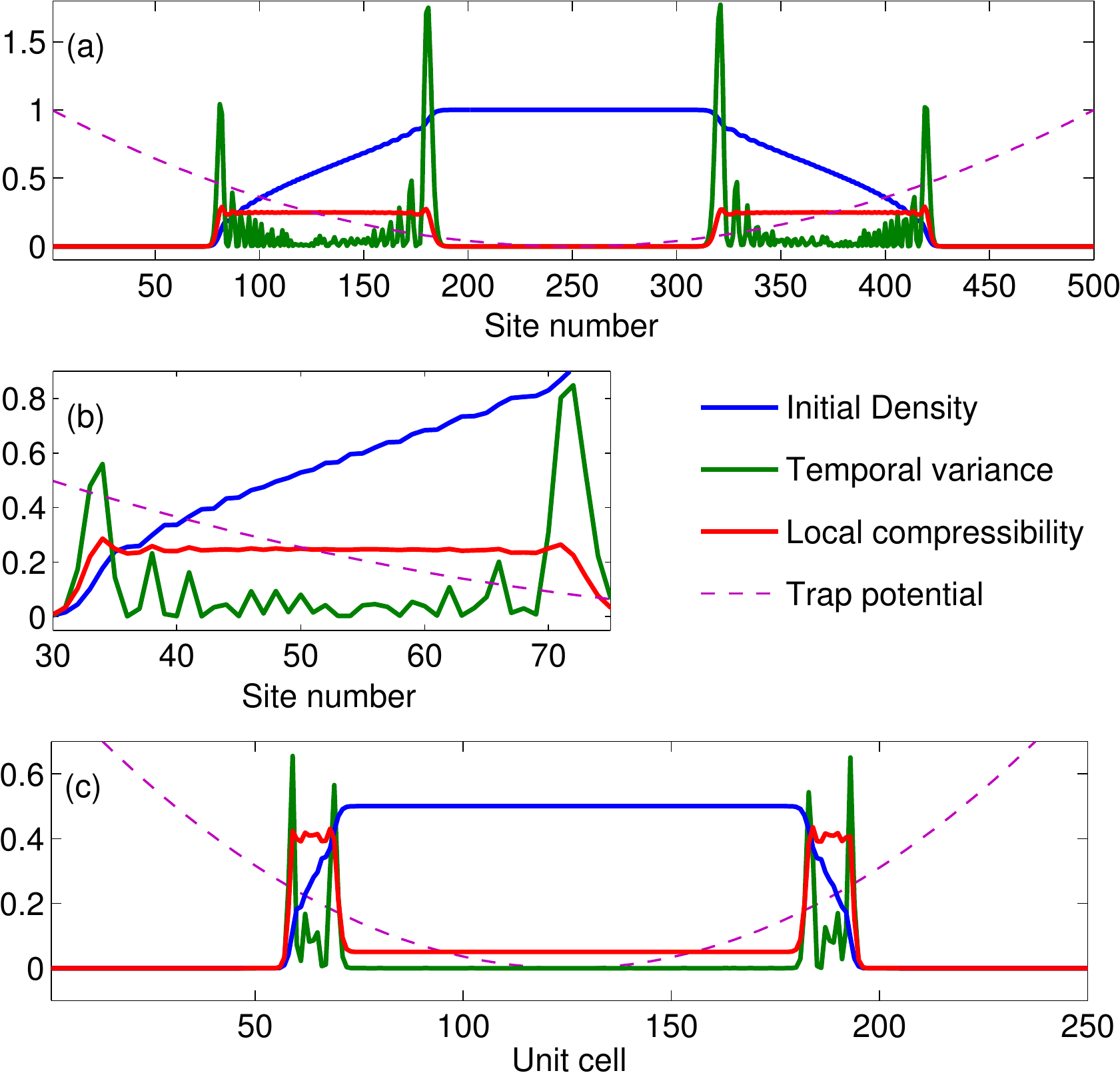} 
\par\end{centering}

\vspace{0.2cm}

\begin{centering}
\includegraphics[clip,width=8cm]{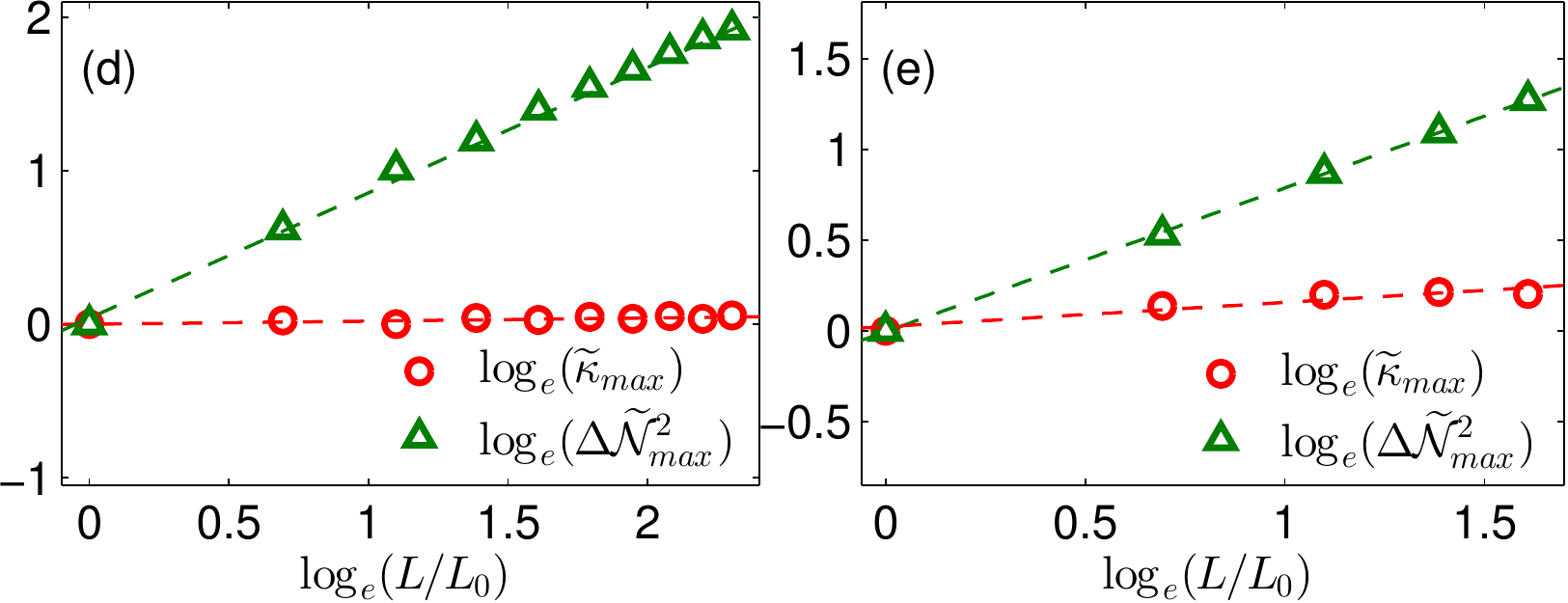} 
\par\end{centering}

\protect\caption{(a) ``Wedding cake'' site occupation profile of hard-core bosons
in a one-dimensional harmonic trap described by Eq.~\eqref{eq:H_stag}
with $V_{0}=0$. The system consists of $L=500$ sites and $N=250$.
The Hamiltonian parameters are $\lambda=10$, $\epsilon=0.2$, $\delta\lambda=L^{-2}$
($J=1$ throughout). The phase boundaries between the Mott plateau
located at the trap center and the adjacent superfluid regions can
be detected by the conventional local compressibility $\kappa_{i}$
(red) and by the temporal variance of the site occupations $\Delta\mathsf{n}_{i}^{2}$
(green) introduced in this work. (b) A closer look at the superfluid
region for the system shown in (a) reveals temporal variance peaks
at the interface between the superfluid and the Mott insulator. (c)
Unit cell average site occupancy in the presence of a staggered potential
Eq.~\eqref{eq:H_stag}. This is a system with $L=500$, $N=150$,
and parameters $\lambda=10$, $\epsilon=0.2$, $\delta\lambda=1/L^{2}$,
$V_{0}=1.5$. (d) and (e) Finite-size scaling of the maximum temporal
variance of the site occupations and of the compressibility vs $L$.
We find $\Delta\mathsf{n}_{\mathrm{max}}^{2}\propto L^{0.83}$ and
$\kappa_{\mathrm{max}}\propto L^{0.05}$ for $V_{0}=0$ (e) and and
$\Delta\mathsf{n}_{\mathrm{max}}^{2}\propto L^{0.80}$ and $\kappa_{\mathrm{max}}\propto L^{0.14}$
for $V_{0}=1.5$. All quantities in the plots are made dimensionless
by dividing by their values when $L_{0}=50$, i.e., $\widetilde{\kappa}_{\mathrm{max}}=\kappa_{\mathrm{max}}(L)/\kappa_{\mathrm{max}}(L_{0})$
and $\Delta\widetilde{\mathsf{n}}_{\mathrm{max}}^{2}=\Delta\mathsf{n}_{\mathrm{max}}^{2}(L)/\Delta\mathsf{n}_{\mathrm{max}}^{2}(L_{0})$.
From \protect\cite{yeshwanth_small_2014}.
\label{fig:quasi-free}}
\end{figure}

For $V_{0}=0$ results of typical simulations are shown in Fig.~\eqref{fig:quasi-free}
panels a), b). In this case the state in the trap center is approximately
the completely filled state $|1,1,\ldots,1,1\rangle$, whereas at
the boundary it is essentially the empty state $|0,0,\ldots,0,0\rangle$.
Clearly both the density fluctuations $\Delta n_{i}^{2}$ and the
local compressibility are able to distinguish the two phases. Both
quantities are approximately constant and non-zero only in the interface
region {[}Fig.~\eqref{fig:quasi-free} panels a), b){]}. Instead
the temporal fluctuations of the site occupation $\Delta\mathsf{n}_{i}^{2}$
show strong fluctuations in the interface region. A closer look at
the density profile in the interface region {[}Fig.~\eqref{fig:quasi-free}
panel b){]} reveals that the density evolves in a stepwise fashion.
Small subregions of constant density are intertwined with pseudo-critical
regions which are properly spotted by the presence of a large temporal
variance. The local compressibility instead is hardly able to resolve
such fine details. Moreover the maxima of the temporal variance diverge
much more rapidly than those of the compressibility. A scaling analysis
reveals that both quantities follow a power law with the following
exponents: $\Delta\mathsf{n}_{\mathrm{max}}^{2}\propto L^{0.83}$,
whereas for the compressibility one finds $\kappa_{\mathrm{max}}\propto L^{0.05}$.
This means that the temporal variance provides a stronger signal as
opposed to the local compressibility. 

Similar results are observed in presence of a nonzero staggerization.
The only caveat is that for $V_{0}\neq0$ one must consider a unit
cell consisting of two neighboring sites {[}see Fig.~\eqref{fig:quasi-free}
panel c){]}. In the trap center the state is approximately $|1,0,1\ldots,1,0\rangle$
separated by an approximately empty state at the boundaries. Once
again both the temporal variance and the local compressibility are
able to distinguish the phases. The finite-size scaling of the maximum
of both quantities, reveals that $\Delta\mathsf{n}_{\mathrm{max}}^{2}\propto L^{0.80}$
and $\kappa_{\mathrm{max}}\propto L^{0.14}$ {[}see Fig.~\ref{fig:quasi-free}(e){]},
respectively, meaning that the temporal variance offers better detectability.

The proposed approached is clearly not limited to integrable models.
We then considered a system consisting of hard-core bosons with nearest
and next-nearest interactions (a $J$-$V$-$V'$ model) in the presence
of a harmonic trap, described by the Hamiltonian 
\begin{align}
\hat{H} & =\sum_{i=1}^{L-1}\left[-J(\hat{b}_{i}^{\dagger}\hat{b}_{i+1}+\hc)+V\left(\hat{n}_{i}-\frac{1}{2}\right)\left(\hat{n}_{i+1}-\frac{1}{2}\right)\right.\nonumber \\
 & \left.+V'\left(\hat{n}_{i}-\frac{1}{2}\right)\left(\hat{n}_{i+2}-\frac{1}{2}\right)\right]+\lambda\sum_{i=1}^{L}i^{2}\hat{n}_{i}.\label{eq:J-V-V1}
\end{align}
Note that, in order to maximize the size of insulating and superfluid
domains, only one half of what would be the harmonic trap is considered
in Eq.~(\ref{eq:J-V-V1}).

In the absence of a trap, the phase diagram of Hamiltonian \eqref{eq:J-V-V1}
has been studied using the density matrix renormalization group technique
\cite{mishra_phase_2011}. The competition between nearest-neighbor
and next-nearest-neighbor interactions generates four phases: two
charge-density-wave insulator phases, a superfluid (Luttinger-liquid)
phase, and a bond-ordered phase. In the presence of a trap, and for
a suitable choice of the parameters, the same four phases can be observed.
We focus our analysis on a parameter regime where the system exhibits
a charge density wave of type one (CDW-I) in the center of the trap,
which is surrounded by a superfluid phase. In this CDW-I phase the
state is approximately $|1,0,1\ldots,1,0\rangle$ as we have seen
previously for model \eqref{eq:H_stag} with $V_{0}\neq0$. However
the CDW-I phase here is not due to the presence of a translationally
symmetry breaking term $V_{0}$ but is stabilized by the presence
of interactions. There are two other phases that have larger unit
cells, consisting of 4 sites for CDW-II and 3 sites for bond-order.
The CDW-I phase is the best suited for our purposes because we are
able to observe several unit cells that exhibit its expected properties.

\begin{figure}
\begin{centering}
\includegraphics[clip,width=8cm]{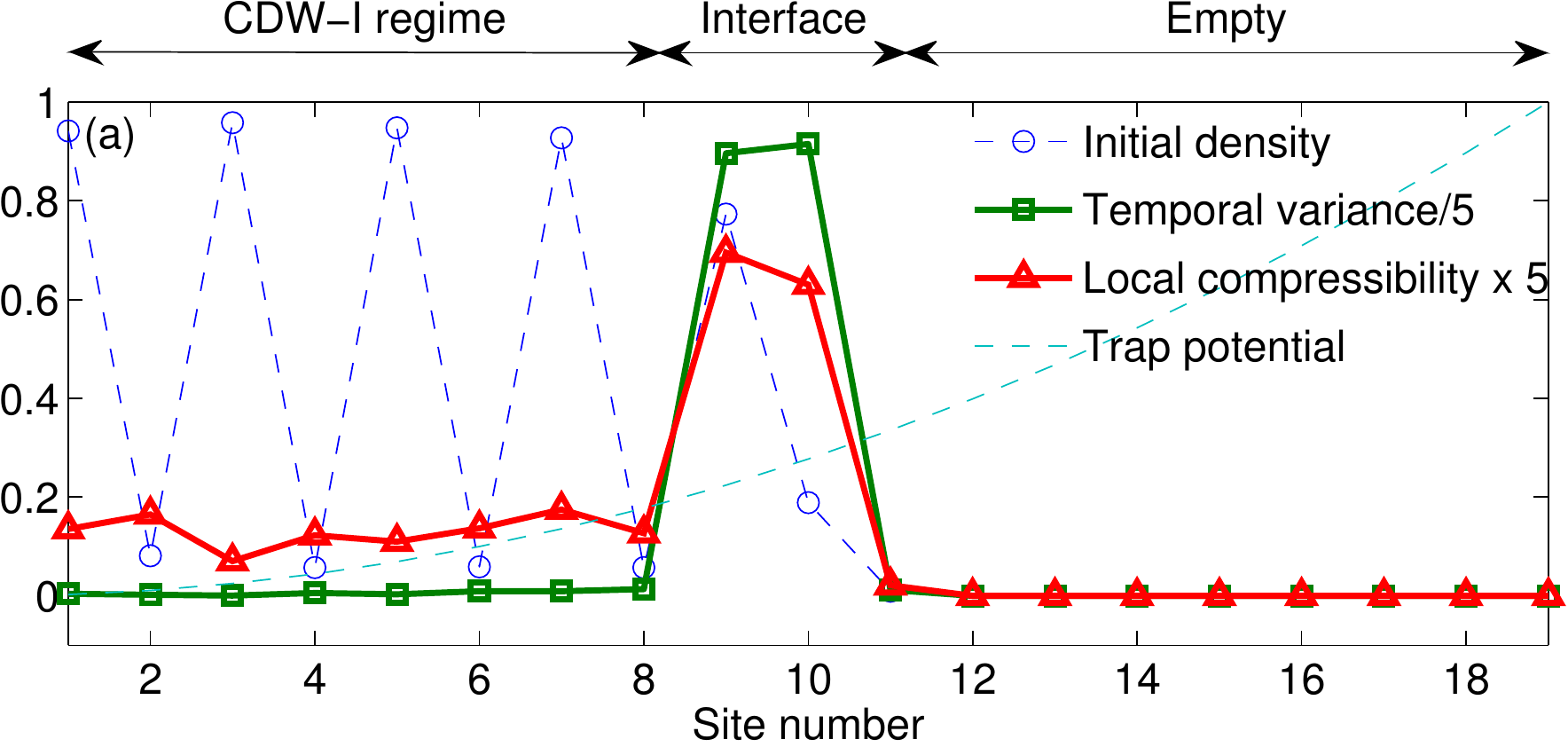} 
\par\end{centering}

\protect\caption{Spatial profile of the temporal variance of the site occupations $\Delta\mathsf{n}_{i}^{2}$
and of the local compressibility $\kappa_{i}$ for the model in Eq.~(\ref{eq:J-V-V1}).
We initialize the system with 19 sites and 5 particles in the ground
state with parameters $J=1$, $V=8.0$, $V'=0.5$ and $\lambda=0.1225$.
The quench is performed by changing the trap potential from $\lambda$
to $\lambda+\delta\lambda$ with $\delta\lambda=0.0061$. From \protect\cite{yeshwanth_small_2014}.
\label{fig:non-integrable}}
\end{figure}

In Fig.~\ref{fig:non-integrable}(a), we show results for a site-occupation
profile exhibiting a CDW-I plateau surrounded by a small superfluid
domain. In the same figure one can see that, at the edge of the CDW-I
plateau, the local compressibility $\kappa_{i}$ exhibits a much weaker
signal than the temporal fluctuations $\Delta\mathsf{n}_{i}^{2}$.
(Note that we used multiplicative factors to enhance $\kappa_{i}$
and reduce $\Delta\mathsf{n}_{i}^{2}$ so that both measures can appear
on the same scale). Also, notice that $\kappa_{i}$ does not vanish
in the CDW-I plateau, which exhibits nonzero site occupation fluctuations.
Since calculations for larger systems are prohibitively large, a finite-size
scaling analysis of the observables is not possible here. Nonetheless,
from Fig.~\ref{fig:non-integrable}(a), it is evident that the temporal
variance is a better indicator of the interface between domains than
the local compressibility. In fact, compared to the integrable systems
considered in the preceding section, the advantage of using $\Delta\mathcal{N}_{i}^{2}$
over $\kappa_{i}$ to identify interfaces between domains is enhanced,
especially taking into account the small system sizes considered here.

\begin{figure}
\begin{centering}
\includegraphics[clip,width=8cm]{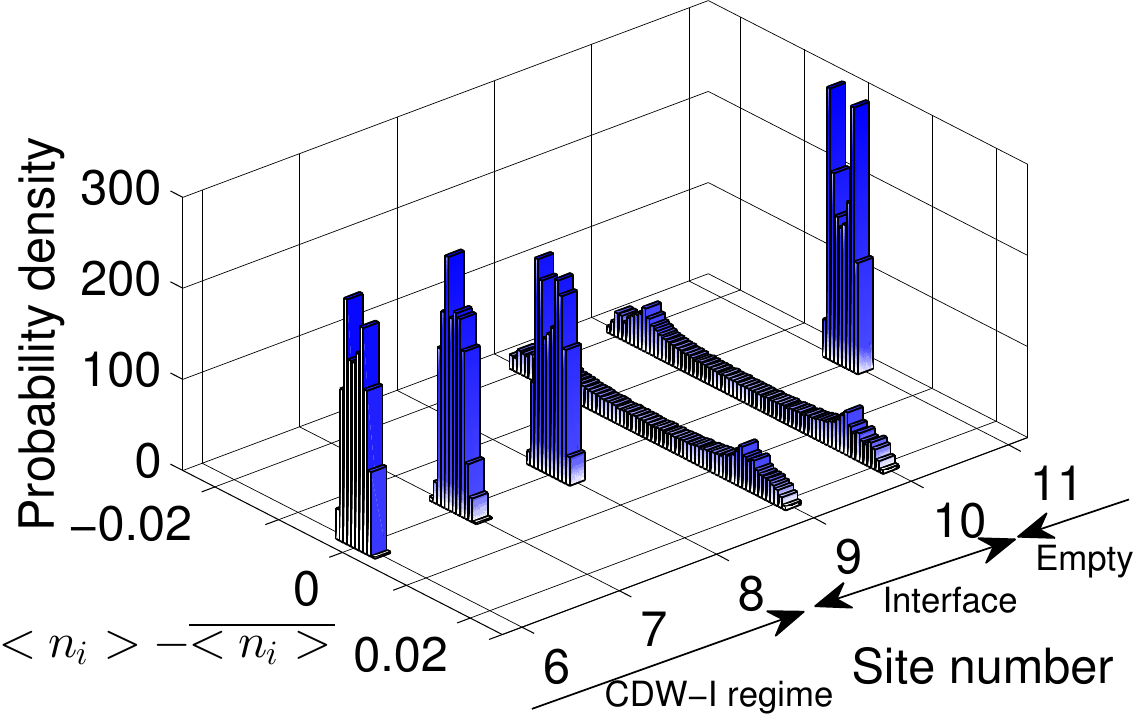} 
\par\end{centering}

\vspace{0.5cm}

\begin{centering}
\includegraphics[clip,width=8cm]{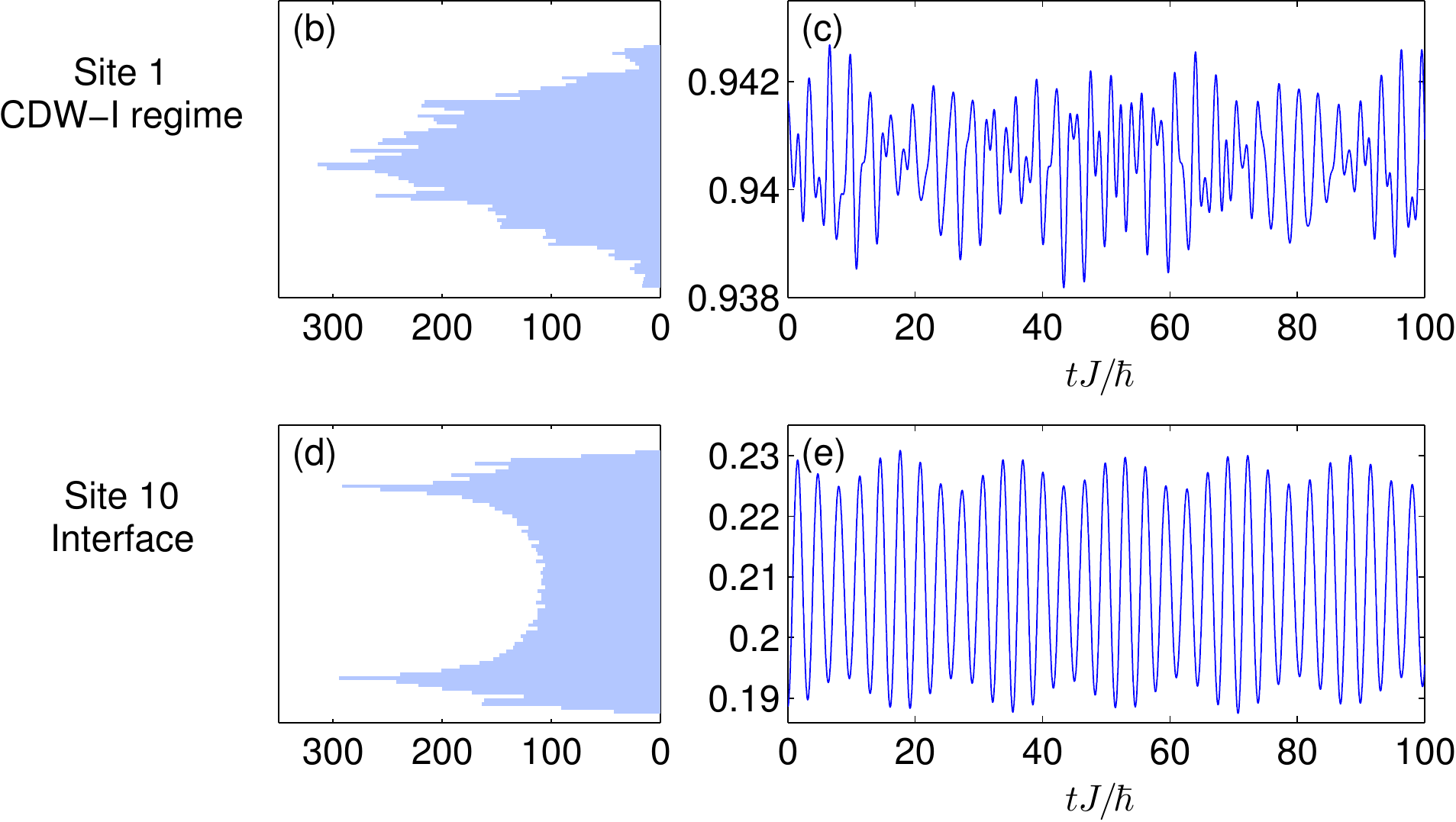} 
\par\end{centering}

\protect\caption{(a) Distributions of the site occupations $\mathsf{n}_{i}(t)$ at
sites near the interface between the CDW-I and the superfluid phase.
The quench amplitude $\delta\lambda$ is the same as for Fig.~\ref{fig:non-integrable}.
(b) and (d) Distribution function of the site occupation at a site
deep in the CDW-I regime (site $i=1$) and at a site at the edge of
the CDW-I domain (site $i=10$), respectively. (c) and (e) Time dependence
of $\mathsf{n}_{i}(t)$ corresponding to (b) and (d), respectively.
The data are obtained by sampling each $\mathsf{n}_{i}(t)$ at $N=4\times10^{4}$
random times uniformly distributed in $\left[0,T\right]$ with $T=40\hbar/J$.
From \protect\cite{yeshwanth_small_2014}. 
\label{fig:full_distributions}}
\end{figure}

One can also go beyond the second moment analysis presented so far
and examine the full probability distribution $P_{i}(x)$ of the random
variable $\mathsf{n}_{i}(t)$ equipped with the time average measure
$\overline{\bullet}$. Based on the results for homogeneous systems
\cite{campos_venuti_universality_2010,campos_venuti_universal_2014},
we expect $P_{i}(x)$ to be a single peaked, approximately Gaussian,
narrow distribution for sites $i$ deep in the (gapped) insulating
regime. On the contrary, $P_{i}(x)$ is predicted to be a double peaked
distribution with a relatively large variance for (critical) interface
sites $i$. In a limiting, somewhat simplified case, $P_{i}(x)$ can
be approximated by a two parameter distribution $P_{i}(x)=1/\left(\pi\sqrt{2\Delta\mathsf{n}_{i}^{2}-(x-\overline{\mathsf{n}_{i}})^{2}}\right)$
\cite{campos_venuti_universality_2010}.

In Fig.~\ref{fig:full_distributions}(a), we show the distribution
$P_{i}(x)$ for sites near the interface separating the insulating
and superfluid regions. For sites $i$ deep in the insulating region
{[}Fig.~\ref{fig:full_distributions}(b){]}, the site occupations
fluctuate about one unique central value, resulting in a singly-peaked
distribution function. This signifies measure concentration, indicating
local equilibration in the finite system considered here {[}Fig.~\ref{fig:full_distributions}(c){]}.
In contrast, as one moves closer to the interface {[}Fig.~\ref{fig:full_distributions}(d){]},
the distribution starts developing two peaks corresponding to a bistability
characteristic of phase boundaries. This breakdown of measure concentration
indicates the breakdown of local equilibration {[}Fig.~\ref{fig:full_distributions}(e){]}
and can thus be used as a witness for spatial phase separation.

\section{The measurement problem}

So far the analysis has been mostly theoretical. We will now try to
address a bit more concretely the problem of determining the temporal
fluctuations $\Delta\mathsf{A}^{2}$ from experimental data. 

So far we assumed, quite naturally, that we can determine, for various
times $t_{j}$, the expectation value $\mathsf{A}(t_{j})=\langle A(t_{j})\rangle$
exactly. This requires to prepare the system many times with the same
initial state and perform say, $N_{s}$ measurements at the \emph{same}
time $t_{j}$ to obtain the $\mathsf{A}(t_{j})$ with sufficient precision
(in principle $N_{s}$ may depend on $j$ but we won't need this generalization
here). Moreover the process will be repeated $N_{d}$ times at different
times, to obtain $\mathsf{A}(t_{1}),\ldots,\mathsf{A}(t_{N_{d}})$.
Since typically the variables are compactly supported (i.e.~$\langle A(t_{j})\rangle$
takes values in a compact set), one can use the Chernoff bound to
deduce that the empirical mean converge to the actual mean exponentially
fast in the number of measurement $N_{s}$. For example, consider
the case where $A$ is a Fermionic number operator at site $i$: $A=c_{i}^{\dagger}c_{i}$.
Denote with $X_{n}$ the result of the $n$-th measurement of $A$
(always after the same preparation time $t_{j}$). In this case $X_{n}$
are Bernoulli trials (i.e.~$X_{n}$ takes only two values, 0,1).
Denoting the empirical mean with $Z_{N_{s}}=N_{s}^{-1}\sum_{n=1}^{N_{s}}X_{n}$
and calling $\mu=\mathsf{E}[Z_{N_{s}}]$ ($\mathsf{E}[\bullet]$ denotes
the expectation value), the Chernoff's bound states that, for $\delta\in(0,1]$,
\begin{equation}
\mathrm{Prob}\left(Z_{N_{s}}<(1-\delta)\mu\right)\le e^{-N_{s}\delta^{2}/2},
\end{equation}
(a similar inequality exist to bound $Z_{N_{s}}$ from below). In
other words, the error one does in estimating $\mu$ is exponentially
small in the number of measurements $N_{s}$. However this may not
be the best strategy to obtain the temporal variance $\Delta\mathsf{A}^{2}$.
In order to design better strategies we must look deeper into the
measurement problem in our out-of-equilibrium setting (the system
is prepared in state $\rho_{0}$ at time $t=0$ and let evolve unitarily
thereafter). Let us indicate with $X_{j}^{n}$ the result of the $n$-th
measurement of $A$ performed at time $t_{j}$. Differently from the
equilibrium case, the variables $X_{j}^{n}$ at different times, are
still independent but are not identically distributed. In the language
of statistics what we would like to build is a \emph{consistent estimator}
of the temporal variance $\Delta\mathsf{A}^{2}$. A consistent estimator
is a method to obtain a given quantity with the property that, as
the number of data point increases, the estimator converges in probability
to the actual parameter we are trying to estimate. In our case the
data points are the variables $X_{j}^{n}$. We now show that it is
possible to estimate $\Delta\mathsf{A}^{2}$ taking $N_{s}$ as small
as $2$. Suppose that we perform two measurements of $A$ at the same
time $t_{j}$, which we denote with $A_{j}^{1},A_{j}^{2}$. The label
$j$ runs form $1$ to $N_{d}$, and we are performing a total of
$N_{s}N_{d}=2N_{d}$ measurements. The following quantity can be shown
to be a consistent estimator of $\Delta\mathsf{A}^{2}$: 
\begin{equation}
s_{2}=\frac{1}{N_{d}}\sum_{j=1}^{N_{d}}A_{j}^{1}A_{j}^{2}-\frac{1}{N_{d}^{2}}\sum_{i,j}A_{i}^{1}A_{j}^{2}.
\end{equation}
Indeed, taking the expectation value, one obtains 
\begin{equation}
\mathsf{E}[s_{2}]=\frac{1}{N_{d}}\sum_{j=1}^{N_{d}}\langle A(t_{j})\rangle^{2}-\Big(\frac{1}{N_{d}}\sum_{j=1}^{N_{d}}\langle A(t_{j})\rangle\Big)^{2}.
\end{equation}
Sampling $t_{j}$ uniformly in $[0,T]$ the above quantity converges
to $\Delta\mathsf{A}^{2}$ as $N_{d}\to\infty$. Alternatively, denoting
with $\mathsf{T}$ the uniform time average over all the different
times $t_{j}$, one has $\mathsf{T}[\mathsf{E}[s_{2}]]=\Delta\mathsf{A}^{2}+O\left(N_{d}^{-1}\right)$,
i.e.~$s_{2}$ is unbiased up to an error $O\left(N_{d}^{-1}\right)$.
Similarly one can show that $\mathrm{\mathsf{var}}[s_{2}]=O\left(N_{d}^{-1}\right)$,
implying the consistency of $s_{2}$. 

In order to minimize the experimental cost, it would be important
to find the most efficient estimators $S(\Delta\mathsf{A}^{2})$ for
the temporal variance. An estimator is efficient if it is unbiased
and if the Cramer-Rao bound \cite{braunstein_statistical_1994},
$\mathrm{\mathsf{var}}[S(\Delta\mathsf{A}^{2})]\ge1/I_{\Delta\mathsf{A}^{2}}$
($I_{\Delta\mathsf{A}^{2}}$ is the Fisher information) is attained.
Clearly further investigations are necessary in this direction, but
these preliminary results indicate that the analysis of the temporal
variance may be an efficient tool to characterize critical properties
with unprecedented details.

\section{Conclusions}

In this review we have described some basic properties of the temporal
fluctuations in isolated, out-of-equilibrium systems. In this setting
a quantum system is initialized in a given state and then let evolve
unitarily undisturbed thereafter. As a consequence, quantum expectation
values of observables become oscillating functions. A great deal of
physical properties are encoded in such temporal fluctuations. In
the general case temporal fluctuations of physical observables are
exponentially small in the system volume. This is encouraging as it
allows to define an average, equilibrium state, with exponential accuracy.
This result is however violated in a few important cases. First of
all, this result does not hold for integrable (quasi-free) systems.
In integrable systems, instead, temporal fluctuations are exponentially
larger and scale with the system's volume. This allows to use temporal
fluctuations to study proximity to integrable points. Temporal fluctuations
can also be completely characterized in a small quench experiment.
In this setting the system is initialized in the ground state of a
given Hamiltonian and then driven out-of-equilibrium by applying a
sudden, small perturbation. Temporal fluctuations can then be used
to characterize the underlying, unperturbed system. If the unperturbed
system is at a regular point of the phase diagram, temporal fluctuations
of generic extensive observables become Gaussian. On the contrary,
close to a quantum critical point, temporal fluctuations acquire a
universal bistable distribution which depends on a single critical
exponent. The results presented here, indicate that temporal fluctuations
may be used in experiments as a tool for probing criticality or integrability
of isolated quantum systems.

What has been left out? The theory of temporal fluctuations presented
in this review parallels, in a way, the theory of quantum fluctuations
of systems at equilibrium \cite{patashinskii_fluctuation_1979}.
In that case one knows that distributions of general observables are
Gaussian, universal, bimodal, for gapped, critical, and symmetry broken
phases respectively. In a similar fashion we have been able to single
out distinctive regimes where the general form of the temporal distribution
can be predicted and the size of the fluctuations estimated. Several
aspects deserve future investigations on the hand of this analogy.
First of all one may ask if other distributions exist in particular
regimes. Furthermore, one may consider temporal autocorrelation functions
of observables $\psi(s)=\overline{\mathsf{A}(t)\mathsf{A}(t+s)}$
which are the analog of the correlations function of equilibrium statistical
mechanics. What kind of informations can be obtained from its study?
More generally, what properties of the asymptotic equilibrium state
\textgreek{r} ¯ can be inferred from the study of temporal fluctuations?
This dynamical setting has much more freedom and complexity than the
equilibrium case and therefore new questions arise. For example, one
may ask how does the equilibration pattern change for slow as opposed
to sudden quenches or what is the effect of quenching from one phase
to another one. The use of temporal fluctuations as a, conceptual
as well as practical, tool has just started to be investigated.

\section*{Acknowledgments}

The author would like to acknowledge the ARO MURI grant W911NF-11-
1-0268 for partial support.
\bibliographystyle{ws-procs9x6}
\bibliography{review_fluctuations,quench_traps_recovered}

\end{document}